\documentclass[%
aps,
reprint,
superscriptaddress,
 amsmath,amssymb,
pra,
]{revtex4-2}

\usepackage{textcomp}
\usepackage[utf8]{inputenc}
\usepackage{gensymb}
\usepackage{graphicx}
\usepackage{dcolumn}
\usepackage{bm}
\usepackage{mathtools}
\usepackage{epstopdf}
\usepackage{hyperref}
\usepackage{color}
\usepackage{nicefrac}
\usepackage{algorithm}
\usepackage{algpseudocode}

\begin{document}
\title{
Three-body renormalization group limit cycles\\based on unsupervised feature learning
}
\author{Bastian Kaspschak}
\affiliation{Helmholtz-Institut f\"ur Strahlen- und Kernphysik and Bethe Center for Theoretical Physics,
Universit\"at Bonn, D-53115 Bonn, Germany}
\author{Ulf-G. Mei{\ss}ner}
\affiliation{Helmholtz-Institut f\"ur Strahlen- und Kernphysik and Bethe Center for Theoretical Physics,
Universit\"at Bonn, D-53115 Bonn, Germany}
\affiliation{Institute for Advanced Simulation, Institut f\"ur Kernphysik, and J\"ulich Center
  for Hadron Physics, Forschungszentrum J\"ulich, D-52425 Jülich, Germany}
\affiliation{Tbilisi State University, 0186 Tbilisi, Georgia}

\date{\today}

\begin{abstract}
Both the three-body system and the inverse square potential carry a special significance in the
study of renormalization group limit cycles. In this work, we pursue
an exploratory approach and address the question which two-body interactions lead
to limit cycles in the three-body system at low energies, without imposing any restrictions upon 
the scattering length. For this, we train a boosted ensemble of
variational autoencoders, that not only provide a severe dimensionality reduction, but also allow to
generate further synthetic potentials, which is an important prerequisite in order to efficiently
search for limit cycles in low-dimensional latent space. We do so by applying an elitist genetic
algorithm to a population of synthetic potentials that minimizes a specially defined limit-cycle-loss.
The resulting fittest individuals suggest that the inverse square potential is the only two-body
potential that minimizes this limit cycle loss independent of the hyperangle.
\end{abstract}

\maketitle

\section{Introduction}
The interest in renormalization group (RG) limit cycles has steadily increased ever since Wilson
has pointed out in 1971 that coupling constants provided by RG equations for field theories of
strong interactions do not necessarily flow towards a fixed point, see Ref.~\cite{PhysRevD.3.1818}.
Instead, RG equations may also allow coupling constants to approach periodical trajectories
in parameter space in a sense that the same coupling constants $c(r_*)=c(\lambda^n r_*)$ are obtained
by multiplying the $n^\text{th}$ power of some preferred scaling factor $\lambda$ to the short-range
cutoff $r_*$, where $n\in\mathbb{N}$. This close connection to discrete scale invariance and
log-periodic cutoff dependence renders a variety of phenomena in particle and nuclear physics as
well as ultracold atoms to be attributable to some kind of RG limit cycle. For instance, while asymptotic
freedom of QCD is associated with an ultraviolet fixed point,
Refs.~\cite{PhysRevLett.91.102002,epelbaum2006more} conjecture the QCD coupling constant to approach
an infrared limit cycle in the three-nucleon system based on tuning the up and down quark masses
and, thereby, further increasing the  magnitudes of the $np$ spin-singlet and spin-triplet
scattering lengths $a_{^{1}\!S_0}=-23.8$~fm and $a_{^{3}\!S_1}=5.4$~fm, which are already large compared
to the spin-triplet effective range $r_0=1.8$~fm (or the pion Compton wavelength, $\lambda_\pi = 1.4\,$fm).
At the critical quark masses the deuteron can be shown to have a vanishing binding energy, while
the triton gains an infinite number of excited states. 
These findings are supported by Ref.~\cite{1434-601X}, 
where triton energies in the critical region are computed up to next-to-next-to-leading order in pionless EFT.

In this context, it is especially worth mentioning the Efimov effect, which has originally been
formulated for a system of three identical bosons attracting each other in the S-wave channel by
one short-ranged two-body interaction, see Refs.~\cite{EFIMOV1970563,osti_4068792}, and has
experienced numerous generalizations to other systems like nucleons, see
Ref.~\cite{doi:10.1146/annurev.nucl.012809.104439}, or macromolecules, e.g. the three-stranded DNA,
see Ref.~\cite{PhysRevLett.110.028105}. In its original formulation it states that particles may
enter three-body bound states even before the potential is strong enough to allow dimers
(two-body bound states) to form. If the interaction becomes resonant, that is if the S-wave
scattering length approaches the unitary limit $a_0\to\pm\infty$, the three-body spectrum becomes
an infinite geometric series
\begin{equation}
\frac{E_3^{(n)}}{E_3^{(n+1)}}=\exp\left(\frac{2\pi}{s_0}\right) \label{eq:three-body-spectrum}
\end{equation}
with the transcendental number $s_0=1.00624$, while the two-body spectrum only consists of a
single zero-energy bound state, see Ref.~\cite{osti_4068792}. The Efimov effect can be explained by
an RG limit cycle with preferred scaling factor $\lambda=\exp(\pi/s_0)= 22.69438$.
Ref.~\cite{PhysRevLett.82.463} provides an alternative approach to the Efimov effect based on
effective field theory and recovers a similar value $s_0\approx 1.0064$. In 2005, first experimental
evidence on the Efimov effect was found in an ultracold gas of caesium atoms by magnetically
tuning scattering lengths based on Feshbach resonances, see Ref.~\cite{kraemer2006evidence}.

It is important to note that the same geometric three-body spectrum as in Eq.~\eqref{eq:three-body-spectrum}
manifests itself independently of the short-range behavior of the two-body potential. This universal property
originates in the fact, that due to $a_0\to\pm\infty$, an effective long-range $1/R^2$ 
behavior in terms of the hyperradius $R$ emerges, see Ref.~\cite{osti_4068792}. Using a suitable
separation ansatz, treating the long-range sector reduces to solving the radial Schr\"odinger equation
for the inverse square potential. Regarding their coupling constants, this implies the same RG flow as
one obtains from directly renormalizing the quantum mechanical $1/r^2$ potential.
The inverse square potential, again, is known to exhibit an RG limit cycle in the two-body sector: 
Ref.~\cite{HAMMER2006306} discusses its renormalization in
momentum space, while Ref.~\cite{PhysRevA.70.052111} compares two different renormalization schemes
in coordinate space: On the one hand, potential well renormalization yields infinitely many branches
of continuous coupling constants, whereas delta-shell renormalization on the other hand
provides one unique coupling constant with infinitely many, log-periodic discontinuities. 
In experiment, the inverse square potential can be reproduced by neutral atoms interacting with
a charged wire, see Ref.~\cite{PhysRevLett.81.737}.
Ref.~\cite{PhysRevLett.108.213202} analyzes the three-body sprectrum of three identical bosons
that pairwisely interact via an actual inverse square potential. In contrast to the Efimov effect,
the inverse square two-body potentials do not need to be resonant and an approximate, but not
exact long-range $1/r^2$ behavior of the resulting three-body potential arises directly from
construction. Most interestingly, an infinite, approximately geometric series of three-body
bound states is shown to exist slightly below the critical strength required to form dimers.
A similar three-body spectrum is later found for a system of three identical fermions, which is
quite intriguing, as the classical Efimov effect requires the pairwise interactions to take
place in the S-wave channel. Due to antisymmetrization, however, this does not hold for the three fermions.

As follows from the detailed explanations in Ref.~\cite{Frank:1971xx}, 
singular potentials are strictly used as
inputs in classical RG analyses and merely specify starting points of RG flow in parameter space.
For a system of three identical bosons, this motivates us to change the paradigm: 
We pursue an exploratory approach that consists of searching
among singular, discretized and finitely-ranged two-body potentials for interactions leading to
RG limit cycles in the three-body sector. Similar to Ref.~\cite{PhysRevLett.108.213202}, we do not
impose any requirements on the S-wave scattering length as the Efimov effect does. Since this system
is only considered at low energies, the low-energy Faddeev equation, see Ref.~\cite{BRAATEN2006259},
applies and can be solved using a generalization of the classical transfer matrix method,
see Ref.~\cite{62122}, for hyperspherical coordinates. Delta-shell regularization is then applied
to the resulting Faddeev wavefunctions. Successively increasing the short-range cutoff provides
detailed information about the RG flow of the corresponding coupling constant. At this point,
a coupling constant exhibiting a log-periodic cutoff dependence, that is discrete scale symmetry
with some preferred scaling factor, indicates an RG limit cycle. As a measure for
log-periodicity we introduce the limit-cycle-loss (LCLoss) on the search space: It is constructed
in such a way that it decreases the closer the RG flow of some coupling constant approaches
an RG limit cycle. Consequently, an exact RG limit cycle is indicated by a vanishing LCLoss.\\

In a machine learning (ML) context, each step of the considered discretized potentials can be
understood as an input feature. A finer discretization, which is required to acquire reliable
approximations of smooth potentials, implies a higher dimensional feature space. Unsupervised feature
learning appears to be a promising approach to gather equally expressive features on much
lower-dimensional vector spaces, see Ref.~\cite{10.1109/TPAMI.2013.50}: We decide to unsupervisedly
train a boosted ensemble of convolutional variational autoencoders (VAEs) to reconstruct two-body
potentials of a targetless training set specially set up for this pretext task. In fact, we benefit
in two ways from this procedure: Firstly, the mentioned ensemble is able to encode high-dimensional
feature vectors to low-dimensional latent vectors, containing only the most distincive information on
the original potential, and vice-versa. Thereby, it provides a severe dimensionality reduction and
allows to relegate our search for RG limit cycles to the much lower-dimensional latent space. Secondly,
it allows to generate infinitely many synthetic potentials, satisfying the feature distributions
inherent to the training set, directly from latent space. The results of the pretext task are a
key ingredient for the downstream task, where the actual search is performed. Here, we apply 
elitist genetic algorithms (GAs) motivated by Goldbergs's Simple Genetic Algorithm,
see Ref.~\cite{goldberg1989genetic}, to several independent populations of synthetic potentials
in parallel, drawn from a multivariate standard distribution in latent space.

This paper is organized as follows: At first, Sec.~\ref{sec:the-three-body-system} briefly recapitulates 
hyperspherical coordinates and the low-energy Faddeev equation. For locally constant
potentials we demonstrate that a simple separation ansatz in terms of a hyperradial and a hyperangular
wavefunction suffices to cover all possible solutions. Accordingly, a generalized transfer matrix method
allows to construct solutions for piecewise-constant potentials. The resulting zero-energy Faddeev
wavefunctions are then used to formulate a matching condition for the delta-shell regularization of the
given two-body potential, which yields the corresponding coupling constant for any hyperangle and
cutoff hyperradius. Proofs of commutation relations, eigenvalue equations and the lengthy computation
of limits, that are necessary to comprehend the results of this section, can be found in
Appendices~\ref{sec:commutator}, \ref{sec:eigenvaluei} and \ref{sec:logarithmichyperradialderivatives}.
Sec.~\ref{sec:generate-potentials} introduces a scaling operation on potentials which causes all
features to be of a similar order of magnitude and, thereby, to be more suitable for ML tasks.
Distinguishing between the short-range and long-range regime, it provides a detailed explanation
how training and test sets consisting only of scaled potentials are generated.
Sec.~\ref{sec:self-supervised-search} motivates the necessity of dimenisonality reduction and to
divide our exploratory approach into a pretext task and a downstream task. In the pretext task we
train a boosted ensemble of convolutional VAEs to reconstruct the scaled potentials from the training set.
Due to boosting, there is a hierarchical order among the individual members of the ensemble, which
is inherited by the encoded potential. This consideration leads to the concept of latent curves from
which synthetic potentials are generated in the downstream task. After having introduced the LCLoss,
we may, therefore, rather understand it as a function mapping a latent curve to some non-negative
number. This, finally, allows to settle on an elitist GA that is applied to fifty populations drawn
from a multivariate standard distribution in latent space. At the end of each GA, we extract the fittest
individual, that is the latent curve with the lowest LCLoss and compare the results in Sec.~\ref{sec:results}.
We end with some outlook on further related investigations.

\section{The three-body system}
\label{sec:the-three-body-system}
\subsection{Equation of motion}
While the two-body problem is classically treated in spherical coordinates, a formulation in
hyperspherical coordinates is advantageous for approaching systems of three particles, as demonstrated
in Ref.~\cite{NIELSEN2001373}. Together with few assumptions on the potential, angular momentum,
and total energy, this leads to the low-energy Faddeev equation, an integro-differential equation
of motion for the three-body system in the low-energy regime.
\subsubsection{Hyperspherical coordinates}
Let all interactions within the three-body system at hand be governed by one spherically-symmetric and
finitely-ranged two-body potential $V$, such that the total potential $V_\text{tot}$ is represented by the sum
\begin{equation}
\begin{aligned}
&V_\text{tot} = V(|\bm{r}_1-\bm{r}_2|) + V(|\bm{r}_2-\bm{r}_3|) + V(|\bm{r}_3-\bm{r}_1|).
\label{eq:totalpotential}
\end{aligned}
\end{equation}
In the following, any distance is  given in units of the range $\rho$ of $V$. Due to the translational
and rotational invariance of $V_\text{tot}$ around the center of mass, only four of the originally
nine degrees of freedom $\bm{r}_1,\bm{r}_2,\bm{r}_3$, with $\bm{r}_i$ representing the position vector
of the $i^\text{th}$ particle, remain. Assuming that all three bosons have the same mass, these are
covered by the hyperradius
\begin{equation}
R = \frac{1}{\sqrt{3}}\sqrt{|\bm{r}_i-\bm{r}_j|^2+|\bm{r}_j-\bm{r}_k|^2+|\bm{r}_k-\bm{r}_i|^2}
\label{eq:hyperradius}
\end{equation}
as well as three hyperangles
\begin{equation}
\alpha_k = \mathrm{arcsin}\left(\frac{|\bm{r}_i-\bm{r}_j|}{\sqrt{2}R}\right)
\label{eq:hyperangle}
\end{equation}
with $(i,j,k)$ being any permutation of $(1,2,3)$, see Ref.~\cite{BRAATEN2006259}. Note that $R$ can
take any value between $0$ and $\infty$, whereas each $\alpha_i$ is restricted to the interval
$[0,\, \pi/2]$. The hyperradius can be understood as the root-mean-square of the three pairwise distances.
A large $R$, therefore, indicates any large $|\bm{r}_i-\bm{r}_j|$ in general. In contrast, the
hyperangles are much more configuration-sensitive. For instance, $\alpha_1 = 0$ represents the scenario
in which the distance between particles~$2$ and $3$ is much smaller than the respective distances to
particle~$1$, which is equivalent to the interaction of a single particle with a two-body cluster.
Vice-versa, if $\alpha_1 = \pi/2$, particle~$1$ is much closer to the center of mass than particles~$2$
and $3$. Finally, there is the special case \linebreak
\begin{equation}
\alpha_i = \frac{\pi}{4} = \mathrm{arcsin}\left(\frac{1}{\sqrt{2}}\right) = \mathrm{arcsin}\left(\frac{|\bm{r}_j-\bm{r}_k|}{\sqrt{2}R}\right) 
\end{equation}
from which we deduce $R = |\bm{r}_j-\bm{r}_k|$. If any other hyperangle takes the same value $\alpha_j
= \pi/4$, this implies that the three particles must be equidistant. Since we have $R = |\bm{r}_k-\bm{r}_i|$
and due to the root-mean-square nature of the hyperradius, the only option left for the remaining
distance between particles $i$ and $j$ is
\begin{equation}
R = |\bm{r}_i-\bm{r}_j| = |\bm{r}_j-\bm{r}_k| = |\bm{r}_k-\bm{r}_i|.
\end{equation}

\subsubsection{Low-energy Faddeev equation}
After separating out the center-of-mass motion and transforming from cartesian to hyperspherical coordinates,
the Schr\"odinger equation reduces to the Faddeev equations. While these are classically a system 
of coupled differential
equations, they can be decoupled by projecting onto S-wave states in order to work entirely in the low-energy regime, 
see Ref.~\cite{NIELSEN2001373}. 
The resulting solutions are superpositions
\begin{equation}
\Psi(R,\bm{\alpha})=\sum\limits_{i=1}^3\psi(R,\alpha_i)
\end{equation}
with the Faddeev wavefunction $\psi(R,\alpha)$ satisfying a single integro-differential equation 
referred to as the low-energy Faddeev equation. As shown in Ref.~\cite{BRAATEN2006259}, together with
the operators
\begin{equation} 
T_R=\frac{1}{2m}R^{-5/2}\left(-\frac{\partial^2}{\partial R^2}+\frac{15}{4R^2}\right)R^{5/2}
\label{eq:hyperradialkineticenergy}
\end{equation}
and
\begin{equation}
T_\alpha = \frac{1}{2mR^2}\frac{1}{\sin(2\alpha)}\left[-\frac{\partial^2}{\partial \alpha^2}-4\right]\sin(2\alpha)
\end{equation}
for the hyperradial and, respectively, hyperangular kinetic energies, this equation is given by
\begin{widetext}
\begin{align}
(T_R+T_\alpha-E)\psi(R,\alpha)=-V(\sqrt{2}R\sin\alpha)\left[ \psi(R,\alpha) + \frac{4}{\sqrt{3}}
\int_{\left|\pi/3-\alpha\right|}^{\pi/2-\left| \pi/6-\alpha\right|} \! \mathrm{d}\alpha^\prime \, \psi(R,\alpha^\prime)
\frac{\sin(2\alpha^\prime)}{\sin(2\alpha)}\right].
\label{eq:lowenergyfaddeev}
\end{align}
\end{widetext}
From a computational perspective, Eq.~\eqref{eq:lowenergyfaddeev} provides an efficient
approach to solve the three-body problem in the low-energy regime.

\subsection{Low-energy wavefunctions}
The following analysis is based on working with finitely-ranged and piecewise-constant two-body potentials.
In order to solve the low-energy Faddeev equation for this class of potentials, we first need to
obtain local solutions at specific potential steps. Using a generalized transfer matrix method for
hyperspherical coordinates to gather all relevant boundary conditions, we then connect these individual
solutions smoothly with each other.

\subsubsection{Separation ansatz for local solutions}
The special case of piecewise-constant two-body potentials $V$ entering Eq.~\eqref{eq:totalpotential}
considerably simplifies the solution procedure. Therefore, we will not perform a hyperspherical
expansion as in Ref.~\cite{BRAATEN2006259}, where the Faddeev wavefunction $\psi$ is decomposed into its
individual channel contributions. Instead, we use the separation ansatz
\begin{equation}
\psi(R,\alpha)=\frac{f(R)}{R^{5/2}}\frac{\phi(\alpha)}{\sin(2\alpha)}
\label{eq:separationansatz}
\end{equation}
which exploits the assumption ${V(\sqrt{2}R\sin(\alpha))=u/2m}$ with $u={\rm const}.$ for some
hyperradius $R$ and hyperangle~$\alpha$ to an even greater extent. Inserting Eq.~\eqref{eq:separationansatz}
as well as the given potential into the low-energy Faddeev equation Eq.~\eqref{eq:lowenergyfaddeev}
allows us to move all hyperangular dependencies to the right-hand side, which defines a hyperradial
function $\gamma$:\\
\vspace{-0.5cm}
\begin{equation}
\begin{aligned}
\frac{\gamma(R)}{2m}&=\frac{R^{5/2}}{f(R)}\left[R^2T_R-\left(E-\frac{u}{2m}\right)\right]\frac{f(R)}{R^{5/2}} \\&= -\frac{\sin(2\alpha)}{\phi(\alpha)}R^2T_\alpha\frac{\phi(\alpha)}{\sin(2\alpha)}\\&\ \ \ -\frac{1}{2m}\frac{4uR^2}{\sqrt{3}\phi(\alpha)}\int_{\left|\pi/3-\alpha\right|}^{\pi/2-\left|\pi/6-\alpha\right|} \! \mathrm{d}\alpha^\prime \, \phi(\alpha^\prime).
\label{eq:separation}
\end{aligned}
\end{equation}
From here, we can deduce two equations of motion that are connected via the expression $\gamma(R)/2m$
on the left-hand side: One equation governing the hyperangular and the other one governing the
hyperradial sector.

\subsubsection{Solving the hyperangular sector}
We first solve the hyperangular sector. The equation of motion for the hyperangular wavefunction $\phi$,
\begin{equation}
\begin{aligned}
&\left[\frac{\partial^2}{\partial \alpha^2}+(4-\gamma(R))\right]\phi(\alpha) \\&-\frac{4uR^2}{\sqrt{3}}
\int_{\left|\pi/3-\alpha\right|}^{\pi/2-\left|\pi/6-\alpha\right|} \! \mathrm{d}\alpha^\prime \, \phi(\alpha^\prime) = 0,
\label{eq:hyperangulareom}
\end{aligned}
\end{equation}
is obtained by rearranging the terms on the right-hand side of Eq.~\eqref{eq:separation} and the
contribution of $\gamma$ on the left-hand side. Note that the hyperradius enters only as a parameter.
Similar to the low-energy Faddeev equation, Eq.~\eqref{eq:hyperangulareom} is also a homogenous,
integro-differential equation. Its shape satisfies 
\begin{equation}
\left[\mathcal{D}(R)-\frac{4uR^2}{\sqrt{3}}\ \mathcal{I}\right]\phi(\alpha)=0,
\label{eq:hyperangulareomsimple}
\end{equation}
where $\mathcal{D}(R)$ and $\mathcal{I}$ denote a differential and, respectively, an integral
operator that are defined by their action on the hyperangular wavefunction $\phi(\alpha)$:
\begin{align}
\mathcal{D}(R)\phi(\alpha) &= \left[\frac{\partial^2}{\partial \alpha^2}+(4-\gamma(R))\right]\phi(\alpha),
\label{eq:operatord}\\
\mathcal{I}\phi(\alpha)  &= \int_{\left|\pi/3-\alpha\right|}^{\pi/2-\left|\pi/6-\alpha\right|} \! \mathrm{d}\alpha^\prime
\, \phi(\alpha^\prime).\label{eq:operatori}
\end{align}
The solution to Eq.~\eqref{eq:hyperangulareomsimple} is much less complicated than it may appear on
first glance. This is because the two operators $\mathcal{D}(R)$ and $\mathcal{I}$ can be easily
shown to commute (see App.~\ref{sec:commutator}), which implies the existence of simultaneous
eigenstates. This observation allows us to search for eigenstates of the individual operators and
subsequently adapt $\gamma$ to the corresponding eigenvalues, such that Eq.~\eqref{eq:hyperangulareom}
is fulfilled.

Due to its fairly simple structure, we start with the eigenvalue equation for the operator $\mathcal{D}(R)$,
\begin{equation}
\left[\frac{\partial^2}{\partial \alpha^2}+(4-\gamma(R))\right]\phi(\alpha) = g \phi(\alpha).
\label{eq:eigenvalueeqd}
\end{equation}
The general solution to Eq.~\eqref{eq:eigenvalueeqd} is given by
\begin{equation}
\begin{aligned}
\phi(\alpha) &= A\ \sin\left(\sqrt{4-\gamma(R)-g}\ \alpha\right) \\&\ \ \ + B\ \cos\left(\sqrt{4-\gamma(R)-g}\
\alpha\right).
\label{eq:generald}
\end{aligned}
\end{equation}
We have to consider the denominator of the separation ansatz in Eq.~\eqref{eq:separationansatz}: In
order to keep the Faddeev wavefunction $\psi$ integrable, the hyperangular wavefunction $\phi$ needs
to vanish simultaneously to the expression $\sin(2\alpha)$ in the denominator. Since hyperangles $\alpha$
are restricted to $0\leq\alpha\leq\pi/2$, this defines exactly two boundary conditions for
Eq.~\eqref{eq:generald},
\begin{equation}
\phi(0)=0, \hspace{1cm} \phi(\pi/2)=0
\end{equation}
Obviously, the coefficient $B$ must vanish to satisfy the first boundary condition. The second boundary
condition restricts the eigenvalues to a discrete set and establishes a connection to the hyperradial
function $\gamma$,
\begin{equation}
g_n(R) = 4(1-n^2)-\gamma(R)
\label{eq:eigenvalued}
\end{equation}
with $n=1,2,3,\ldots\,$. Note that choosing $n=0$ is forbidden as this yields $g_0(R)=4-\gamma(R)$:
When inserting $g_0$ into Eq.~\eqref{eq:generald}, we see that this leads to the trivial solution
$\phi(\alpha)=0$. The eigenstates corresponding to the eigenvalues $g_n$ then turn out to be the simple modes
\begin{equation}
\phi_n(\alpha)=\sin(2n\alpha).
\label{eq:eigenstatesd}
\end{equation}
It can be easily checked that these eigenstates are a complete orthogonal system, which agrees with
the hermitecity of $\mathcal{D}(R)$. Since we did not perform a hyperspherical expansion, the index
$n$ does not label the individual channels as in Ref.~\cite{BRAATEN2006259}. Instead, considering
the eigenstates $\phi_n$ once more, $n$ can rather be understood as a node index, similar to the
radial quantum number in the hydrogen atom.

In App.~\ref{sec:eigenvaluei} we examine how the integral operator $\mathcal{I}$ acts on the
eigenstates $\phi_n$ from Eq.~\eqref{eq:eigenstatesd}. In fact, for each node index $n=1,2,3,\ldots$
the eigenstate $\phi_n$ of $\mathcal{D}(R)$ also turns out to be an eigenstate of $\mathcal{I}$,
\begin{equation}
\mathcal{I}\phi_n(\alpha)=\frac{1}{n}\sin\left(\frac{2\pi n}{3}\right)\phi_n(\alpha).
\label{eq:eigenvaluei}
\end{equation}
Most interestingly, the corresponding eigenvalues ${i_n=\sin\left(2\pi n/3\right)/n}$ appear to be
non-degenerate with one severe exception. If the node index $n=3n^\prime$ is a multiple of three,
we observe $i_{3n^\prime}=0$ with $n^\prime=1,2,3,\ldots\,$. This not only shows that the eigenvalue
$i_{n}=0$ is degenerate, but its eigenspace and, therefore, the kernel of $\mathcal{I}$ is infinitely
dimensional.

Now we insert the simultaneous eigenstates $\phi_n$ into Eq.~\eqref{eq:hyperangulareomsimple} and
substitute the operators $\mathcal{D}(R)$ and $\mathcal{I}$ by the respective eigenvalues given in
Eqs.~\eqref{eq:eigenvalued} and \eqref{eq:eigenvaluei},
\begin{equation}
\begin{aligned}
&\left[\mathcal{D}(R)-\frac{4uR^2}{\sqrt{3}}\  \mathcal{I}\right]\phi_n(\alpha)\\&\ \hspace{1cm}
=\left[g_n(R)-\frac{4uR^2}{\sqrt{3}n}\sin\left(\frac{2\pi n}{3}\right)\right]\phi_n(\alpha)\\
&\ \hspace{1cm} \overset{!}{=}0,
\label{eq:eqforgamma}
\end{aligned}
\end{equation}\newpage

\noindent
As Eq.~\eqref{eq:eqforgamma} must hold for any node index $n$, this finally defines a discrete family
of hyperradial functions
\begin{equation}
\gamma_n(R)=4\left[1-n^2-\frac{uR^2}{\sqrt{3}n}\sin\left(\frac{2\pi n}{3}\right)\right]~.
\label{eq:gammadef}
\end{equation}\\

\subsubsection{Solving the hyperradial sector}
The equation of motion for the hyperradial sector of the low-energy Faddeev equation follows
from the first equality in Eq.~\eqref{eq:separation},
\begin{equation}
\left[2mR^2T_R-\left(k^2-u\right)-\gamma_n(R)\right]\frac{f_{k,n}(R)}{R^{5/2}}=0,
\end{equation}
where $k^2=2mE$ is the square of the total momentum. Since it is clear that the hyperradial wavefunctions
must depend on $k$ as well as on the node index $n$ due to its connection to the hyperangular sector,
we have included pertinent indices provisionally. We use the expression from Eq.~\eqref{eq:gammadef}
for the family of hyperradial functions $\gamma_n$ we gathered by solving the hyperangular sector.
After several steps of term rearrangement we arrive at the equation
\begin{equation}
R^2\frac{\mathrm{d}^2f}{\mathrm{d}R^2} + [k_n(u)R]^2f_{k,n}(R)-\left(n^2-\frac{1}{4}\right)f(R)=0.
\label{eq:hyperradialeom}
\end{equation}
The $k_n(v)$ can be understood as modified momenta and are defined by
\begin{equation}
k_n(u) = \xi(k)\sqrt{k^2-\left[ 1+\frac{4}{\sqrt{3}n}\sin\left(\frac{2\pi n}{3}\right) \right]u}
\label{eq:momenta-knu}
\end{equation}
on the entire complex plane. The binary function
\begin{equation}
\xi(k) = \begin{cases}
 +1 & \mathrm{for~Re}(k) \geq 0 \\
-1 & \mathrm{for~Re}(k) < 0
\end{cases}
\end{equation}
ensures that the total momentum is correctly reproduced, that is $k_n(0)=k$, if the potential
vanishes locally or, respectively, $k_{3n}(u)=k$ if the node index is a multiple of three on
the entire complex plane. Note that $k_n(u)^2/2m$ corresponds to the kinetic energy in the limit
$n\to\infty$ of an infinitely large node index.
Eq.~\eqref{eq:hyperradialeom} is related to the Bessel equation and is solved by the linear combination
\begin{equation}
\begin{aligned}
f_{k,n}(R) &=A_{k,n}\sqrt{R}\ J_n\left[k_n(u) R\right] \\ &\ \hspace{1cm} +B_{k,n}\sqrt{R}\ Y_n\left[k_n(u)
 R\right]~,
\label{eq:hyperradialwave}
\end{aligned}
\end{equation}
with $J_n$ and $Y_n$ denoting the $n^\text{th}$ Bessel function of first and second kind, respectively.
Finally, we can combine the hyperangular wavefunctions $\phi_n$ from Eq.~\eqref{eq:eigenstatesd}
and the hyperradial wavefunction \eqref{eq:hyperradialwave} according to the separation
ansatz Eq.~\eqref{eq:separationansatz} and obtain an expression for the Faddeev wavefunction,
\begin{widetext}
\begin{equation}
\psi_{k,n}(R,\alpha) = \left\{A_{k,n}\ J_n\left[k_n(u) R\right] +B_{k,n}\ Y_n\left[k_n(u) R\right]\right\}
\frac{\sin(2n\alpha)}{\sin(2\alpha)R^2}
\label{eq:faddeevwavefunction}
\end{equation}
\end{widetext}

\subsubsection{Transfer matrix method for hyperspherical coordinates}
\begin{figure}[t]
\includegraphics[width=\columnwidth]{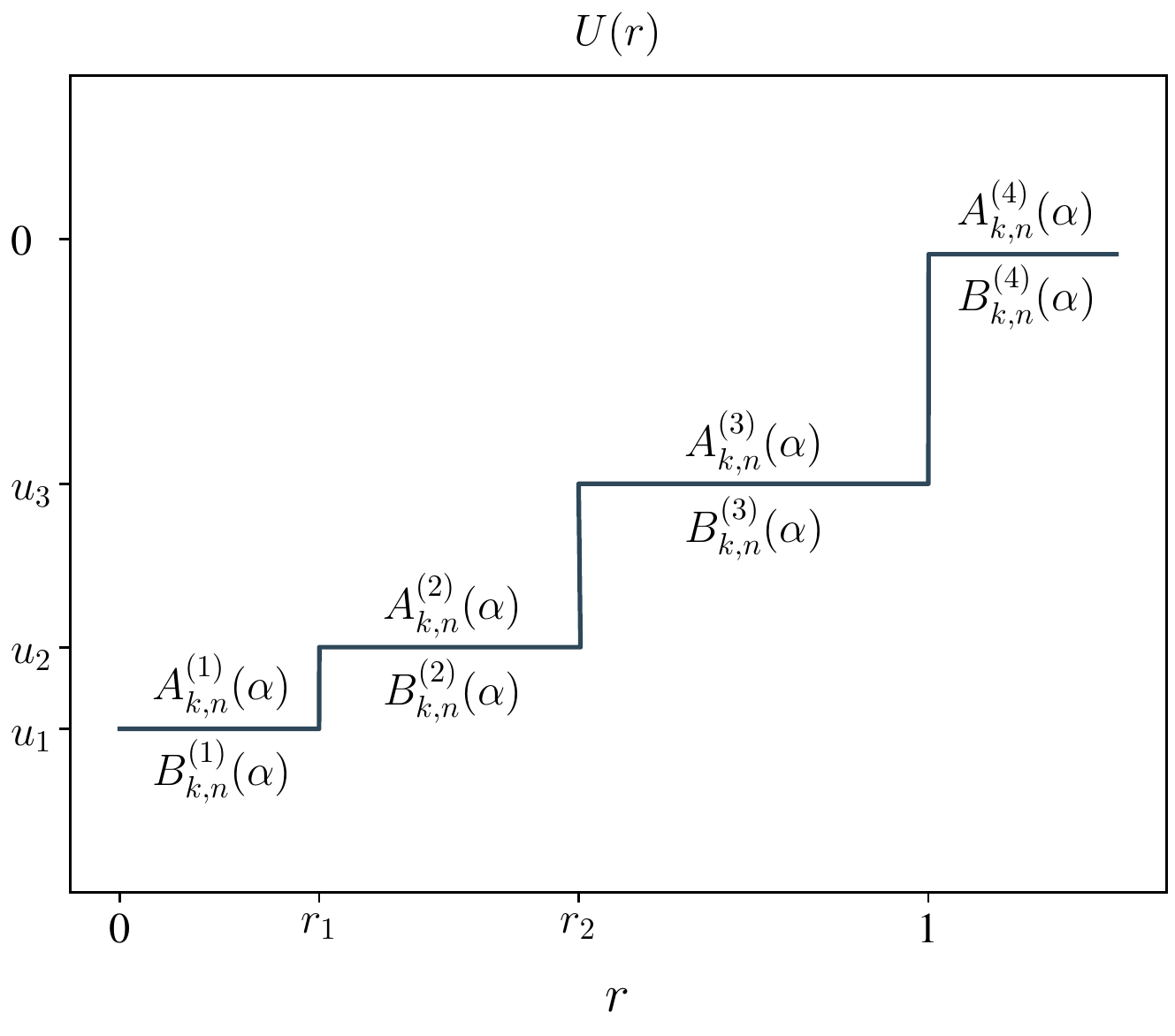}
\caption{Piecewise-constant two-body potential with three transition radii $r_1$, $r_2$ and $r_3$.
Due to $r=\sqrt{2}R\sin(\alpha)$, we can adopt the classical transfer matrix method to determine Bessel
coefficients for hyperspherical coordinates. Keeping the hyperangle $\alpha$ fixed, we apply transfer
matrices for adjacent hyperradii $R$. As a consequence, $\alpha$ enters both Bessel coefficients as
a parameter.}
\label{fig:sampled_potential}
\end{figure}
Up to a normalization, the Bessel coefficients $A_{k,n}$ and $B_{k,n}$ in Eq.~\eqref{eq:faddeevwavefunction}
can be arbitrarily chosen. However, Eq.~\eqref{eq:faddeevwavefunction} assumes the two-body potential
$V$ to be constant in the vicinity of $r=\sqrt{2}R\sin(\alpha)$ for some hyperradius $R$ and
hyperangle $\alpha$. A piece-wise constant two-body potential $V$ is constant between any adjacent
transition radii $0\leq r_{i-1}<r_i\leq 1$, that is
\begin{equation}
V(r_{i-1}\leq r<r_i)=\frac{u_i}{2m},
\label{eq:transradii}
\end{equation}
where the origin $r_{0}=0$ serves effectively as the zeroth transition radius. In contrast, the final
transition radius is always $r_{F}=1$. Radii beyond $r_{F}$, with $F$ being the number of all potential
steps, correspond to radii beyond the range of $V$, which imposes $u_{F+1}=0$.
In the following we will refer to the index $i$ enumerating the steps $u_i$ of the potential as the step
index. The transition radius $r_i$ can be translated into transition hyperradius $R_i$ that also
depends on the hyperangle,
\begin{equation}
R_i(\alpha)=\frac{r_i}{\sqrt{2}\sin(\alpha)}
\label{eq:transhyperradii}
\end{equation}
Eq.~\eqref{eq:transhyperradii} allows to translate Eq.~\eqref{eq:transradii} into hyperspherical
coordinates. An examplary potential with three non-zero transition radii is shown in
Fig.~\ref{fig:sampled_potential}. From now on, we work with dimensionless potentials $U = 2mV$
for which we obtain the hyperspherical segmentation
\begin{equation}
U(\sqrt{2}R\sin(\alpha))=u_i \hspace{0.5cm}\text{if}\ \ R_{i-1}(\alpha) \leq R < R_i(\alpha).
\end{equation}

For some hyperangle~$\alpha$, we know that the Faddeev wavefunction $\psi_{k,n}(R,\alpha)$ must behave
like the solution given in Eq.~\eqref{eq:faddeevwavefunction} between any pair of transition hyperradii
$R_{i-1}(\alpha)$ and $R_i(\alpha)$,
\begin{equation}
\begin{aligned}
\psi^{(i)}_{k,n}(R,\alpha) &= \left\{A_{k,n}^{(i)}(\alpha)\ J_n\left[k_n^{(i)} R\right] \right. \\
&\ \ + \left. B_{k,n}^{(i)}(\alpha)\ Y_n\left[k_n^{(i)} R\right]\right\}\frac{\sin(2n\alpha)}{\sin(2\alpha)R^2}~,
\label{eq:faddeevwavefunction-transfer}
\end{aligned}
\end{equation}
with $k_n^{(i)}=k_n(u_i)$. 
If generalized to hyperspherical coordinates, the transfer matrix method, see Ref.~\cite{62122}, appears to
be quite promising for providing the hyperradial and hyperangular dependences of the Bessel coefficients.
The basic idea of this hyperspherical transfer matrix method is to leverage the continuity 
\begin{equation}
\begin{aligned}
&A_{k,n}^{(i)}(\alpha)\ J_n\left[k_n^{(i)} R_i(\alpha)\right]\\&\ \hspace{0.5cm}+B_{k,n}^{(i)}(\alpha)\
Y_n\left[k_n^{(i)} R_i(\alpha)\right] \\ &= A_{k,n}^{(i+1)}(\alpha)\ J_n\left[k_n^{(i+1)} R_i(\alpha)\right]\\
&\ \hspace{0.5cm}+B_{k,n}^{(i+1)}(\alpha)\ Y_n\left[k_n^{(i+1)} R_i(\alpha)\right]
\end{aligned}
\label{eq:continuity}
\end{equation}
and differentiability 
\begin{equation}
\begin{aligned}
&A_{k,n}^{(i)}(\alpha)k_n^{(i)}\ J_n^\prime\left[k_n^{(i)} R_i(\alpha)\right]\\&\ \hspace{0.5cm}
+B_{k,n}^{(i)}(\alpha)k_n^{(i)}\ Y_n^\prime\left[k_n^{(i)} R_i(\alpha)\right] \\
&= A_{k,n}^{(i+1)}(\alpha)k_n^{(i+1)}\ J_n^\prime\left[k_n^{(i+1)} R_i(\alpha)\right]\\
&\ \hspace{0.5cm}+B_{k,n}^{(i+1)}(\alpha)k_n^{(i+1)}\ Y_n^\prime\left[k_n^{(i+1)} R_i(\alpha)\right]
\end{aligned}
\label{eq:differentiability}
\end{equation}
of the Faddeev wavefunction to formulate two boundary conditions on the Bessel coefficients at each
transition hyperradius $R_i(\alpha)$ of the piecewise-constant potential. Eqs.~\eqref{eq:continuity}
and \eqref{eq:differentiability} form a system of linear equations that can be solved for the Bessel
coefficients
\begin{equation}
\begin{pmatrix} A_{k,n}^{(i+1)}(\alpha) \\ B_{k,n}^{(i+1)}(\alpha) \end{pmatrix} = \frac{\pi}{2}R_i(\alpha)\
\mathrm{M}_{k,n}^{(i)}(U,  \alpha) \begin{pmatrix} A_{k,n}^{(i)}(\alpha) \\ B_{k,n}^{(i)}(\alpha) \end{pmatrix}.
\label{eq:transfer}
\end{equation}
Eq.~\eqref{eq:transfer} can be understood as a single vector equation relating the Bessel coefficients of
the two potential steps around some transition hyperradius $R_i(\alpha)$ for a given hyperangle $\alpha$
via a multiplication with the transfer matrix  
\begin{equation}
\mathrm{T}_{k,n}^{(i)}(V,  \alpha) = \frac{\pi}{2}R_i(\alpha)\ \mathrm{M}_{k,n}^{(i)}(U,  \alpha).
\end{equation}
Using the product
\begin{equation}
\begin{aligned}
&\left\{f, g\right\}_{k,n}^{(i)}(U,  \alpha) \\&= k_n^{(i+1)}f_{n}\left[k_n^{(i)} R_i(\alpha)\right]g_{n-1}
\left[k_n^{(i+1)} R_i(\alpha)\right]\\ &\ \hspace{0.3cm}-k_n^{(i)}f_{n-1}\left[k_n^{(i)} R_i(\alpha)\right]g_{n}
\left[k_n^{(i+1)} R_i(\alpha)\right]
\end{aligned}
\end{equation}
for two countable function sequences $\{f_n\}_{n\in\mathbb{N}}$ and $\{g_n\}_{n\in\mathbb{N}}$, the matrices
$\mathrm{M}$ can be shown to satisfy the pattern
\begin{equation}
\mathrm{M} = \begin{pmatrix} +\left\{J, Y\right\} & +\left\{Y, Y\right\} \\ -\left\{J, J\right\}
& -\left\{Y, J\right\} \end{pmatrix}.
\label{eq:matrices}
\end{equation}
In Eq.~\eqref{eq:matrices} we have kept the dependences on the hyperangle $\alpha$, the two-body
potential $U$, the node index $n$ and the step index $i$ implicit for the sake of brevity.

The naive hyperspherical transfer matrix method allows to freely choose Bessel coefficients
$A_{k,n}^{(1)}$ and $B_{k,n}^{(1)}$ for the step index $i=1$ and then apply a chain of the required
transfer matrices to obtain the Faddeev wavefunction at any hyperspherical configutation $(R,\alpha)$.
However, choosing $B_{k,n}^{(1)}=0$ is necessary due to the singularity of the Bessel functions $Y_n$ of
second kind at the origin. We can then choose $A_{k,n}^{(1)}=1$ as the wavefunction can still be
normalized afterwards. This completely determines all remaining coefficients from
Eq.~\eqref{eq:faddeevwavefunction} as
\begin{equation}
A_{k,n}^{(i)}(\alpha) = \begin{pmatrix} 1, & 0 \end{pmatrix}\left( \prod\limits_{j=1}^{i-1}
 T_{k,n}^{(j)}(U, \alpha)\right) \begin{pmatrix} 1 \\ 0 \end{pmatrix}
\end{equation}
and
\begin{equation}
B_{k,n}^{(i)}(\alpha) = \begin{pmatrix} 0, & 1 \end{pmatrix}\left( \prod\limits_{j=1}^{i-1}
T_{k,n}^{(j)}(U, \alpha)\right) \begin{pmatrix} 1 \\ 0 \end{pmatrix}.
\end{equation}

There is a residual factor $1/R^2$ in Eq.~\eqref{eq:faddeevwavefunction}, which on first glance causes
the Faddeev wavefunction to diverge in the origin, even if $B_{k,n}^{(1)}=0$. Fortunately, this does not
break the normalizability of the Faddeev wavefunction due to the following reasons: Firstly,
${\psi_{k,n}^{(1)}(R,\alpha)=J_n[k_n^{(i)}R_i(\alpha)]}$ has at least a first-order root in the origin,
since the smallest allowed node index is $n=1$. This definitely eliminates one factor $1/R$.
Finally, the second $1/R$ factor is eliminated by the Jacobi determinant during a hyperspherical
integration.
\subsection{Delta-shell regularization}
The piecewise-constant two-body potentials that we later work with approximate singular potentials.
These again are known to produce unphysical infinities in the short-range sector, see
Ref.~\cite{PhysRevA.64.042103}. Since we are only interested in the low-energy sector, this issue
can be remedied by a suitable renormalization method. 

The key idea of the Wilsonian renormalization group approach is to eliminate the short-range
degrees of freedom above some UV-cutoff by regularizing the potential and introducing additional
cutoff-dependent couplings. It is essential that the renormalized potential faithfully reproduces
chosen low-energy observables, which defines a matching condition on the new couplings.
\begin{figure}[t]
\includegraphics[width=\columnwidth]{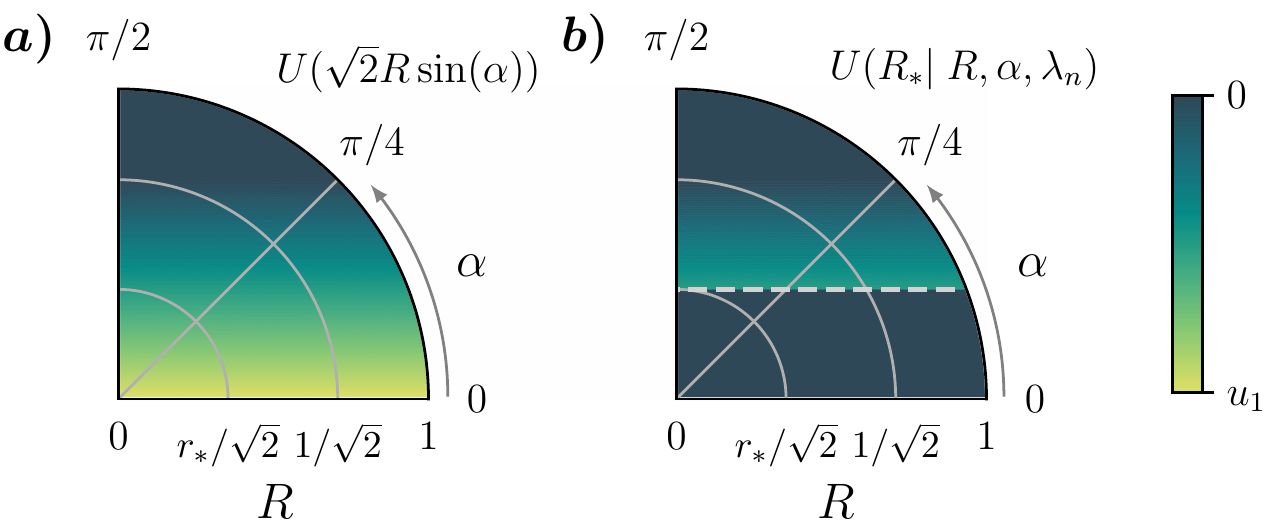}
\caption{Given the  linear potential $V(r)=v_1(r-1)/(2m)$, in Fig.~\ref{fig:delta_polar}a) the
corresponding polar plot using the relation $r=\sqrt{2}R\sin(\alpha)$ is displayed. Note that
all contours are represented by horizontal lines $R(\alpha)=r / (\sqrt{2}\sin(\alpha))$ with fixed $r$.
With this knowledge, a delta-shell regularization with short-range cutoff $r_*$ can be visually
understood as eliminating the potential for all $(R,\alpha)$ below the line $R_*(\alpha)
=r_*/(\sqrt{2}\sin(\alpha))$, as shown in Fig.~\ref{fig:delta_polar}b).}
\label{fig:delta_polar}
\end{figure} 

Delta-shell regularization, which is thoroughly explained in Ref.~\cite{PhysRevA.70.052111}, is a
convenient regularization technique for potentials in configuration space. In contrast to
related techniques like potential well regularization, it provides a unique coupling and the
short-range cutoff is not bounded from below. The standard delta-shell regularization substitutes
a given two-body potential $U$ up to some cutoff radius $r_*$ by a delta-shell potential, such that
the regularized potential reads
\begin{equation}
U(r_*|\ r, H) = \begin{cases}-\displaystyle\frac{H(r_*)}{r_*}\delta(r-r_*) & r<r^* \\
+U(r) & r\geq r_* 
\end{cases}
\label{eq:classicdeltashell}
\end{equation}
with the coupling constant $H(r_*)$.
Since we are about to match the logarithmic derivatives of the hyperradial zero-energy wavefunctions,
we have to formulate Eq.~\eqref{eq:classicdeltashell} in terms of hyperspherical coordinates.
This specific matching condition introduces respective dependences on $R$ and $\alpha$ as well as
a dependence on the node index $n$ to the coupling constant and, lastly, to the renormalized potential.
A very important step is to translate the radial delta distribution to a hyperradial one.
Analogously to the cutoff radius $r_*$ we may define a cutoff hyperradius $R_*=r_*/\sqrt{2}\sin(\alpha)$
for some hyperangle~$\alpha$. As $\alpha$ only runs from $0$ to $\pi/2$, we can be sure that
$\sin(\alpha)\geq 0$. 
For the delta distribution this means that we can extract a factor $\sqrt{2}\sin(\alpha)$ from its argument,
\begin{equation}
\delta(r-r_*)=\frac{1}{\sqrt{2}\sin(\alpha)}\delta(R-R_*).
\label{eq:hyperradialdelta}
\end{equation}
\newpage
Inserting Eq.~\eqref{eq:hyperradialdelta} into Eq.~\eqref{eq:classicdeltashell} and implementing the
above mentioned adjustments then yields the hyperspherical delta-shell potential,
\vfill
\noindent
\begin{equation}
U_n(R_*|\ R, \alpha, H_n)=-\frac{1}{2R_*\sin^2(\alpha)}H_n(R_*,\alpha)\delta(R-R_*) \hspace{0.5cm}
\label{eq:hypersphericaldeltashell}
\end{equation}
\vfill
\noindent
for hyperradii $R<R_*$ enclosed by the delta-shell.\linebreak A delta-shell regularization eliminates
all transition hyperradii $R_{i-1}<R_*$ for any hyperangle $\alpha$ as the regularized
potential $U_n(R_*|\ R, \alpha, H_n)$ uniquely vanishes inside of the delta-shell, see
Fig.~\ref{fig:delta_polar}a) in comparison to Fig.~\ref{fig:delta_polar}b). This makes
$\widehat{R}_1=R_*$ effectively the first transition hyperradius of the regularized potential and
imposes $\widehat{u}_1=0$. All other transition hyperradii remain unchanged due to the limited
range of the delta-shell potential, that is $\widehat{R}_2=R_i$, $\widehat{R}_3=R_{i+1}$, $\ldots$
together with the corresponding potential steps $\widehat{u}_2=u_i$, $\widehat{u}_3=u_{i+1}$, $\ldots$
and so on. 
In contrast to $H(r_*)$ from Eq.~\eqref{eq:classicdeltashell}, we allow the 
coupling constant $H_n(R_*,\alpha)$ in Eq.~\eqref{eq:hypersphericaldeltashell} to be an arbitrary 
function of $R_*$ and $\alpha$. This is clearly a different situation from the standard formulation 
of the Efimov effect, but ensures that logarithmic derivatives of Faddeev wavefunctions 
can be matched at all hyperangles.

\vfill

When it comes to determining $H_n(R_*,\alpha)$, the hyperradial delta distribution invites to
integrate over the low-energy Faddeev equation. It is helpful to simplify the low-energy
Faddeev equation even further by inserting the eigenvalues of the integral operator $\mathcal{I}$
from Eq.~\eqref{eq:eigenvaluei},
\vfill
\begin{equation}
\begin{aligned}
&2m(T_R+T_\alpha-E)\psi_{0,n}(R,\alpha) \\ &= -U_n(R_*|\ R,\alpha,H_n)\left[1+\frac{4}{\sqrt{3}}
\mathcal{I}\right]\psi_{0,n}(R,\alpha)\\
&=-U_n(R_*|\ R,\alpha,H_n)\left[1+\frac{4}{\sqrt{3}n}\sin\left(\frac{2\pi n}{3}\right)\right]
\psi_{0,n}(R,\alpha).
\label{eq:lowenergyfaddeev-deltashell-1}
\end{aligned}
\end{equation}
\vfill
\noindent
Eq.~\eqref{eq:lowenergyfaddeev-deltashell-1} has to be understood as a direct relation between the
coupling constant and the zero-energy Faddeev wavefunction. Proceeding from here, an integradion
over the infinitesimal interval $[R_*-\delta R,R_*+\delta R]$ covering the cutoff hyperradius yields
\vfill
\begin{equation}
\begin{aligned}
&H_n(R_*,\alpha)= -2R_*\sin^2(\alpha)\int_{R_*-\delta R}^{R_*+\delta R} \mathrm{d}R \, U_n(R_*|\ R,\alpha,H_n)\\
&=\frac{4mR_*\sin^2(\alpha)}{1+\frac{4}{\sqrt{3}n}\sin\left(\frac{2\pi n}{3}\right)}\\&\ \ \
\times\int_{R_*-\delta R}^{R_*+\delta R} \mathrm{d}R \, \frac{1}{\psi_{0,n}(R,\alpha)}(T_R+T_\alpha-E)
\psi_{0,n}(R,\alpha).
\label{eq:infinitesimalintegral}
\end{aligned}
\end{equation}
\vfill
\noindent
While the zero-energy Faddeev wavefunction itself, of course, must be continuous, the same does not
hold for its hyperradial first-order derivatives as a consequence of the delta-shell potential.
Although this means that the right-hand side of Eq.~\eqref{eq:infinitesimalintegral} does not necessarily
vanish, most terms of its integrand do become negligible due to integrating over an infinitesimal interval.
In fact, the only non-vanishing contribution comes from the second-order hyperradial derivative of
the corresponding kinetic energy operator, c.f. Eq.~\eqref{eq:hyperradialkineticenergy}, acting on
the hyperradial zero-energy wavefunction,
\vfill
\begin{equation}
\begin{aligned}
&H_n(R_*,\alpha) = -\frac{2R_*\sin^2(\alpha)}{1+\frac{4}{\sqrt{3}n}\sin\left(\frac{2\pi n}{3}\right)}
\int_{R_*-\delta R}^{R_*+\delta R} \mathrm{d}R \, \frac{\widehat{f}_{0,n}^{\prime\prime}(R)}{\widehat{f}_{0,n}(R)}\\
&=\frac{-2R_*\sin^2(\alpha)}{1+\frac{4}{\sqrt{3}n}\sin\left(\frac{2\pi n}{3}\right)}\left[
\lim_{R\to R_*^-}\frac{\widehat{f}_{0,n}^{(1)\prime}(R)}{\widehat{f}_{0,n}^{(1)}(R)}-\lim_{R\to R_*^+}
\frac{\widehat{f}_{0,n}^{(2)\prime}(R)}{\widehat{f}_{0,n}^{(2)}(R)}\right].
\label{eq:difflogderivs}
\end{aligned}
\end{equation}
\vfill
\noindent
Here, $\widehat{f}_{0,n}^{(1)}$ denotes the hyperradial wavefunction inside of the first segment of the
regularized potential. We assume the cutoff hyperradius $R_*$, which serves as the first transition
hyperradius of the regularzed potential, to be in the $i^\text{th}$ segment of the unregularized potential.
This allows to relate the corresponding hyperradial wavefunction segment with each other via
$\widehat{f}_{0,n}^{(2)} = f_{0,n}^{(i)}$.
\vfill

On the right-hand side of Eq.~\eqref{eq:difflogderivs} we see that the coupling constant is completely
determined by matching the hyperradial logarithmic derivatives of the hyperradial zero-energy wavefunction
close to the cutoff hyperradius. The calculation of the hyperradial logarithmic derivatives is
relegated to App.~\ref{sec:logarithmichyperradialderivatives}. With Eq.~\eqref{eq:difflogderivs} in mind,
we need to use the results 

\vfill
\begin{widetext}
\begin{equation}
\lim_{R\to R_*^+}\frac{f_{0,n}^{(i)\prime}(R)}{f_{0,n}^{(i)}(R)}=k_n^{(i)}\frac{A_{0,n}^{(i)}(\alpha)J_{n-1}[k_n^{(i)}R_*]
+B_{0,n}^{(i)}(\alpha)Y_{n-1}[k_n^{(i)}R_*]}{A_{0,n}^{(i)}(\alpha)J_{n}[k_n^{(i)}R_*]+B_{0,n}^{(i)}(\alpha)
Y_{n}[k_n^{(i)}R_*]} + \frac{1-2n}{2R_*}
\end{equation}
and
\begin{equation}
\lim_{R\to R_*^-}
\frac{\widehat{f}_{0,n}^{(1)\prime}(R)}{\widehat{f}_{0,n}^{(1)}(R)}=\frac{1+2n}{2R_*}
\end{equation}
to arrive at
\begin{equation}
H_n(R_*,\alpha)=\frac{2\sin^2(\alpha)}{1+\frac{4}{\sqrt{3}n}\sin\left(\frac{2\pi n}{3}\right)}
\left[2n-k_n R_*\frac{A_{0,n}^{(i)}(\alpha) J_{n-1}[k_n R_*] + B_{0,n}^{(i)}(\alpha) Y_{n-1}[k_n R_*]}
{A_{0,n}^{(i)}(\alpha) J_n[k_n R_*] + B_{0,n}^{(i)}(\alpha) Y_n[k_n R_*]}\right].
\label{eq:lambdacoupling}
\end{equation}
\end{widetext}

\clearpage

\section{Generating scaled, singular two-body potentials}
\label{sec:generate-potentials}
The first step in our search for limit cycles is to decide on a specific class of potentials, which
defines the search space, and to generate corresponding data sets. In Eq.~\eqref{eq:totalpotential}
we have already restricted the total potential to superpositions of two-body potentials. For the sake
of brevity, we refer to potentials as being LC, if their coupling constant $H_n(R_*,\alpha)$,
obtained by a delta-shell regularization, exhibits an RG limit cycle in the three-body sector.
If such a limit cycle, however, does not manifest, we consequently call them non-LC.

In order to apply the hyperspherical transfer matrix method, the generated, attractive
potentials need to be piecewise-constant. Ref.~\cite{PhysRevA.70.052111} suggests to search
especially among singular potentials for LC potentials. With this we are interested on the one hand
in faithfully approximating smooth potentials, e.g. of type $1/r^n$, as well as simultaneously
covering the short-range and long-range regime. These two criteria require us to use numerous
non-equidistant transition radii and turn out to be satisfactorily fulfilled by
\begin{equation}
r_i = \mathrm{e}^{12(i-F)/(F-1)} \label{eq:transition-radii-exp}
\end{equation}
for $i=1,\ldots, F$, with $F=10^3$. These suffice to probe coupling constants $H_n(R_*,\alpha)$
for the notably large range $\log(R_*)\in[-12-\log(\sqrt{2}\sin(\alpha)),-\log(\sqrt{2}\sin(\alpha))]$
of logarithmic cutoff hyperradii.

By fixing the number of transition radii, we have automatically specified each potential to consist
of exactly $F=10^3$ negative segments with values $u_i$. Following machine learning terminology,
we refer to the individual $u_i$ as features. This allows us to understand potentials as vectors
$\bm{U}=(u_1,\ldots,u_F)^\top$ in feature space $\mathbb{R}^F$ and, accordingly, the number $F$
of features to be the feature space dimension.

The problem of naively applying machine learning algorithms to search among singular potentials
is that their features strongly vary in range. Instead, the scaled potentials $\bm{\widetilde{U}}$
that are componentwisely related to the original potentials $\bm{U}$ via
\begin{equation}
\widetilde{u}_i = \frac{1}{8} \log(-u_i). \label{eq:potential-scaling-rel}
\end{equation}
have their features $\widetilde{u}_i$ at a similar order of magnitude and are more suitable
for machine learning.
Note that while all $u_i$ are negative, the scaled features $\widetilde{u}_i$ can vanish and in
general take positive and negative values.

\subsection{Short-range behavior of $\widetilde{U}$}
During the construction of data sets it is necessary to keep physical reasonability in mind.
For instance, oscillations in the short-range sector are not resolvable. 
Regarding its short-range behavior, we therefore construct $\widetilde{\bm{U}}$ to be strictly
decreasing. This motivates us to generate the short-range part separated from the long-range
part as follows:
\begin{minipage}{\linewidth}
\begin{algorithm}[H]
  \caption{Generate Scaled Potential - Short Range}
  \label{alg:potential-short}
   \begin{algorithmic}[1]
	\State \textbf{take} $\log(r)$ from \textbf{uniform distribution} $\mathcal{U}([-8.4, -3.6])$
	\State $\bm{x} \leftarrow (-12, \log(r), 0)^\top$
	\State \textbf{take} $\bm{y}$ from \textbf{uniform distribution} $\mathcal{U}([0, 7.5]^3)$
	\State \textbf{sort} $\bm{y}$ in descending order
	\State $f \leftarrow$ \textbf{quadratic spline interpolation} w.r.t. $\bm{x}$ and $\bm{y}$
	\State $\widetilde{\bm{U}}_\text{short} \leftarrow$ \textbf{empty list}
	\For{$i\in\{1,\ldots, F\}$}
			\State \textbf{append} $f(\log(r_i))$ \textbf{to} $\widetilde{\bm{U}}_\text{short}$
  \EndFor
	\State \textbf{sort} $\widetilde{\bm{U}}_\text{short}$ in descending order
	\State \textbf{apply} \textbf{Savitzky-Golay filter} of order $3$ with window size $601$ to $\widetilde{\bm{U}}_\text{short}$
	\State \textbf{sort} $\widetilde{\bm{U}}_\text{short}$ in descending order
	\For{$i\in\{1,\ldots, F\}$}
			\State $(\widetilde{U}_\text{short})_i\leftarrow(\widetilde{U}_\text{short})_i-(\widetilde{U}_\text{short})_F$
  \EndFor
	\State \textbf{return} $\widetilde{\bm{U}}_\text{short}$ \Comment{short-range part of generated potential}
   \end{algorithmic}
\end{algorithm}
\end{minipage}\\

\noindent
Note that the features $(\widetilde{U}_\text{short})_i$ uniquely determine the short-range behavior
due to imposing the normalization $(\widetilde{U}_\text{short})_F=0$. This is equivalent to the
condition ${U_\text{short}(1)=-1}$ for the unscaled short-range contribution.

\subsection{Long-range behavior of $\widetilde{U}$}
Having generated the short-range part $\widetilde{\bm{U}}_\text{short}$, we are only missing the
long-range part $\widetilde{\bm{U}}_\text{long}$ in order to obtain the scaled potential $\widetilde{\bm{U}}$ as
\begin{equation}
\widetilde{\bm{U}}=\widetilde{\bm{U}}_\text{short}\odot\widetilde{\bm{U}}_\text{long}.
\label{eq:shortlongrangevtilde}
\end{equation} 
Here we use the Hadamard product defined as elementwise multiplication $(\bm{x}\odot\bm{y})_i=x_iy_i$.
In contrast to the short-range sector, resolvability does not impose any technical limitations here,
which is why oscillations may also come at play:
\begin{minipage}{\linewidth}
\begin{algorithm}[H]
  \caption{Generate Scaled Potential - Long Range}
  \label{alg:potential-long}
   \begin{algorithmic}[1]
	\State \textbf{take} $num$ from \textbf{uniform distribution} $\mathcal{U}(\{2,\ldots, 20\})$
	\State $\bm{x} \leftarrow \left(0,1/(num-1),2/(num-1),\ldots, 1\right)^\top\in\mathbb{R}^{num}$
	\State $\bm{y}\leftarrow (1)$
	\For{$i\in\{1,\ldots, num-1\}$} \Comment{uniform random walk}
				\State \textbf{take} $step$ from \textbf{uniform distribution} $\mathcal{U}([-0.5, 0.5])$
				\State \textbf{append} $y_i+step$ \textbf{to} $\bm{y}$
  \EndFor
	\State $f \leftarrow$ \textbf{quadratic spline interpolation} w.r.t. $\bm{x}$ and $\bm{y}$
	\State $\widetilde{\bm{U}}_\text{long} \leftarrow$ \textbf{empty list}
	\For{$i\in\{1,\ldots, F\}$}
				\State \textbf{append} $f(r_i)$ \textbf{to} $\widetilde{\bm{U}}_\text{long}$
  \EndFor
	\State \textbf{apply} \textbf{Savitzky-Golay} filter of order $3$ with window size $101$ to $\widetilde{\bm{U}}_\text{long}$ \textbf{twice} in a row
	\State \textbf{return} $\widetilde{\bm{U}}_\text{long}$ \Comment{long-range part of generated potential}
   \end{algorithmic}
\end{algorithm}
\end{minipage}
\begin{figure}[ht]
\includegraphics[width=0.9\columnwidth]{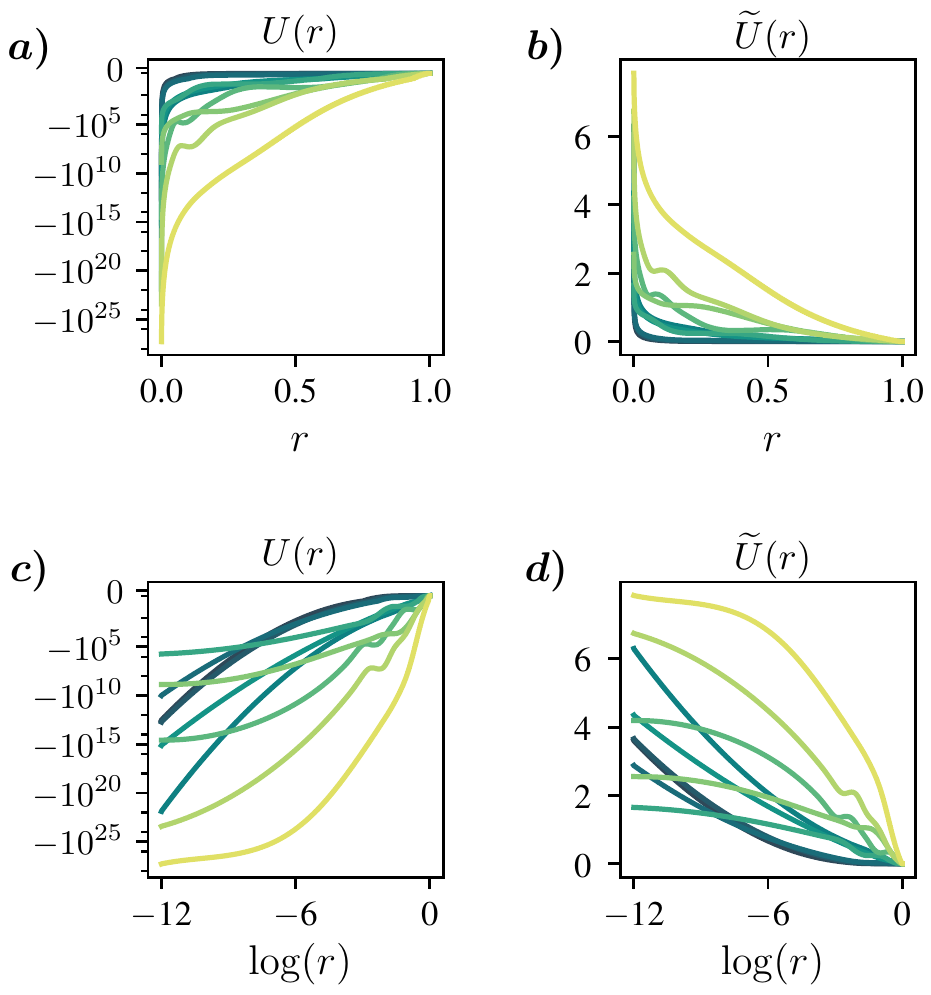}
\caption{Ten randomly generated, singular two-body potentials are plotted over the radius $r$ in
Fig.~\ref{fig:data_raw_scaled}a). Fig.~\ref{fig:data_raw_scaled}b) displays the corresponding
scaled potential $\widetilde{U}$. 
Figs.~\ref{fig:data_raw_scaled}c) and d) plot $U$ and $\widetilde{U}$ over the logarithm of
the radius. As can be seen, potentials constructed this way behave monotonically in the
short-range regime. Vice-versa, oscillations may only occur in the long-range part.}
\label{fig:data_raw_scaled}
\end{figure} 

\subsection{Training and test data sets}
Combining the results of Algorithms~\ref{alg:potential-short} and \ref{alg:potential-long}
according to Eq.~\eqref{eq:shortlongrangevtilde} yields one scaled two-body potential.
Fig.~\ref{fig:data_raw_scaled}b) displays ten of such randomly generated potentials. By construction,
the corresponding unscaled potentials shown in Fig.~\ref{fig:data_raw_scaled}a) are singular in
the origin. Comparing with Figs.~\ref{fig:data_raw_scaled}c) and d), we convince ourselves that
oscillations only appear in the long-range regime, whereas the short-range part behaves monotonous.

In this fashion, we generate $3\times 10^4$ training potentials and $3\times 10^3$ test potentials
forming the training set $X$ and, respectively, the test set $Y$.

\section{Self-supervised search for limit cycles}
\label{sec:self-supervised-search}
Both data sets $X$ and $Y$ generated in Sec.~\ref{sec:generate-potentials} are targetless and only
contain singular, scaled potentials as shown in Fig.~\ref{fig:data_raw_scaled}. Common supervised
machine learning techniques are, therefore, ruled out. Instead, the search for LC potentials
can be understood as an optimization problem that aims at minimizing some loss function directly
processing bare, scaled potentials. This loss function is introduced in Sec.~\ref{sec:limit-cycle-loss}
as the limit cycle loss (LCLoss) and measures how close a potential is to being LC. 
Naively, one could think of this optimization to take place in feature space. Since we do not assume
to have already found LC potentials during the creation of the data sets $X$ and $Y$, an essential part
of this search is to generate new potentials, also called synthetic potentials. These synthetic
potentials neither appear in $X$ or $Y$, but have to satisfy the respective potential distributions.
However, due to the high dimension of the feature space, several problems may occur, that are briefly
referred to as the curse of dimensionality, see Ref.~\cite{661089}. It is clear that not every point
in feature space corresponds to a singular, scaled potential as those in $X$ and $Y$. This makes
generating synthetic potentials a non-trivial task that heavily relies on knowledge about patterns
in both data sets as well as the fundamental distribution of potentials in feature space to
draw new samples from. Finally, for faithfully estimating the mentioned potential distribution,
the data sets with $|X|=3\times 10^4$ and $|Y|=3\times 10^3$ certainly do not contain enough samples.

Fortunately, the feature space dimension $F$ exceeds the number of degrees of freedom in both data
sets by far, which strongly invites to apply suitable dimensionality reduction techniques. Thus,
we also need distinguish between a pretext task and a downstream task. The pretext task finds a
low-dimensional representation containing the most distinctive information on scaled, singular
potentials and, thereby, mitigates the curse of dimensionality. Hereafter, the downstream task
performs the actual search for LC potentials and benefits from the results of the pretext task
in several ways: Firstly, knowing the low-dimensional representation much better allows to estimate
a corresponding low-dimensional potential distribution to draw synthetic potentials from. Secondly,
due to being carried out in a low-dimensional vector space, optimization techniques can be assumed to
converge faster and to yield more robust results. 

Note that the downstream task no longer relies on the data sets $X$ and $Y$, but entirely runs on
more abstract, machine learned features found by the pretext task. Therefore, the described procedure
can most likely be attributed to unsupervised feature learning, see Ref.~\cite{10.1109/TPAMI.2013.50}.
\subsection{Pretext task}
There are numerous dimensionality reduction techniques worth mentioning: The principal component
analysis (PCA), see Ref.~\cite{doi:10.1080/14786440109462720}, is a quite popular method. It projects a
data set onto the low-dimensional vector space $\mathbb{R}^L$ spanned by those $L\in\mathbb{N}$
eigenvectors of its correlation matrix that have the highest eigenvalues. Since the variance of the
data is maximal along these axes, they are associated to the $L$ most distinctive features of the data
set. This makes PCA a rewarding approach for classification and anomaly detection tasks given smaller
and less complex data sets, see Refs~\cite{Valentino_2017, doi:10.1063/1.2945165}. Being a linear
method, it is, however, not suited to extract complex, non-linear patterns.
Ref.~\cite{10.5555/2987061.2987133} discusses the helix problem to explain the severe limitations
of PCA when dealing with non-linear data: The helix problem is a non-linear toy problem in which
points are closely distributed along a helix in $\mathbb{R}^3$. While it is intuitively clear
that there is only one degree of freedom at play, a PCA tends to overestimate the number of required
principal components to be $L=3$ and, therefore, fails to reduce the dimension of the helix-problem.
In addition, Ref.~\cite{10.1109/TPAMI.2013.50} points out, that stacking several PCAs does not
yield more abstract and expressive features, as this sequence is again a linear operation and,
therefore, does not enhance non-linearity.

In the case of singular two-body potentials within an $F=10^3$ dimensional feature space, a non-linear
approach, e.g. via autoencoders is required. An autoencoder $\mathcal{A}=\mathcal{D}\circ\mathcal{E}:
\mathbb{R}^F\to\mathbb{R}^F$ maps the feature space $\mathbb{R}^F$ to itself and is defined as the composition
of an encoder $\mathcal{E}:\mathbb{R}^F\to \mathbb{R}^L$ and a decoder $\mathcal{D}:\mathbb{R}^L\to
\mathbb{R}^F$, see Refs.~\cite{bourlard1988auto,10.5555/2987189.2987190}. Both, the encoder and the decoder
are typically fully-connected (FC), convolutional (CNN) or recurrent neural networks (RNN) that are
non-linearily activated and can be trained via gradient descent. They either map to or, respectively,
map from the latent space $\mathbb{R}^L$, which plays a central role in this concept of non-linear
dimensionality reduction. The application of $\mathcal{E}$ to a point in feature space is called
encoding. Consequently, applying $\mathcal{D}$ to some point in latent space is referred to as
decoding. Having both, an encoder and a decoder at hand, allows us to easily associate an entire
distribution $X$ in feature space with the distribution $\mathcal{E}(X)$ of the corresponding,
encoded data in latent space.

The training objective for autoencoders is to reproduce inputs $\widetilde{\bm{U}}\in X$ as
faithfully as possible. This is accompanied by the so-called reconstruction loss
\begin{equation}
  \mathcal{L}_\text{Rec}^{(\mathcal{A})}(\widetilde{V})=\mathcal{L}(\mathcal{A}(\widetilde{\bm{U}}),
  \widetilde{\bm{U}})
\end{equation}
measuring the deviation of a given input $\widetilde{\bm{U}}$ from its reconstruction
$\mathcal{A}(\widetilde{\bm{U}})$. For $\mathcal{L}$ we decide to use the L1-Loss 
\begin{equation}
\mathcal{L}_1(x,y)=\frac{1}{n}\sum_{i=1}^n|x_i-y_i|\label{eq:l1loss}
\end{equation}
with $x,y\in\mathbb{R}^n$, although other loss functions like the MSELoss $\mathcal{L}_\text{MSE}(x,y)
= \sum_{i=1}^n(x_i-y_i)^2/n$, for instance, would, of course, also be a reasonable choice.
\begin{figure*}[t]
\includegraphics[width=\textwidth]{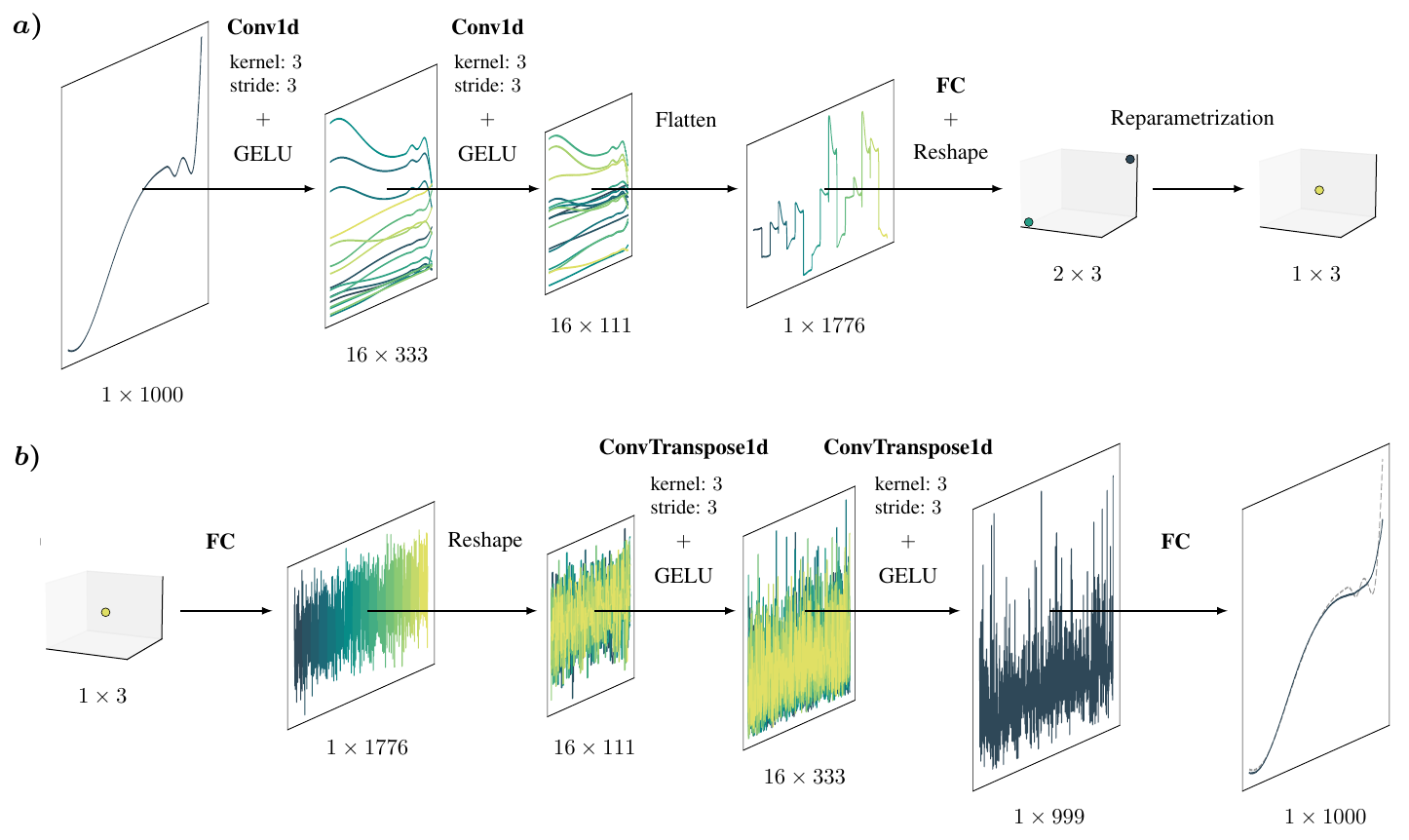}
\caption{Encoder a) and decoder architecture b) of the employed VAEs.
  Since they contain two non-linearily activated
convolutional layers each, we understand the resulting VAEs as deep neural networks. The encoder pipeline
finishes with the reparametrization trick yielding a point in three dimensional latent space. This point
again contains the most essential information about the original input and serves as starting point of
the decoder pipeline for input reconstruction. In fact, the last window in b)
allows to compare the reconstruction (solid line) with the original input (dashed line). Note that the
VAEs have been trained on standardized data, which is why the input shows some standardized
potential $\widehat{\bm{U}}$ with components $\widehat{U}_i=(\widetilde{U}_i-\mu^{(X)}_i)/\sigma^{(X)}_i$
with the elementwise mean $\mu^{(X)}$ and, respectively, standard deviation $\sigma^{(X)}$ of $X$.}
\label{fig:cvae_architecture}
\end{figure*}

Note that in the case of equal feature and latent space dimensions, $L=F$, the reconstruction loss is
minimized whenever the encoder and decoder counter-act each other, such that $\mathcal{A}$ approaches
the identity in feature space. However, if no further sparse coding techniques as described in
Ref.~\cite{pmlr-v2-ranzato07a} are applied, the downstream task hardly benefits from the newly
acquired latent space features. Instead, in order to obtain a significant dimensionality reduction,
the case $L\ll F$ is particularly interesting, as it strictly enforces the encoder to act as a
projection onto the low-dimensional latent space. Vice-versa, the decoder performs as an embedding
from latent space back into the much higher-dimensional feature space. This feature extraction paradigm
is also known as the bottleneck method, see Ref.~\cite{tishby2000information}. On the one hand, such
a bottleneck architecture may impose a severe information loss during encoding. On the other hand,
if the reconstruction loss has been sufficiently minimized during training, this implies that the
encoded potentials $\mathcal{E}(\widetilde{\bm{U}})$ contain the $L$ most distinctive features of the
underlying data set. The decoder, again, has learned to reconstruct the essential behavior of the
original scaled potential $\widetilde{\bm{U}}$.

Note that the training objective hardly influences the distribution $\mathcal{E}(X)$ of encoded potentials
in latent space. During encoding, we cannot eliminate the case that minor changes in feature space
blow up and become large deviations in latent space. Therefore, the encoded distribution in latent
space may not only turn out to be disjoint and multimodal, but the Euclidean metric in latent space
is no faithful similarity measure for the elements of feature space, as well. This severely complicates
exploratory approches based on synthetic potentials. As the downstream task strongly depends on the
latter, we need to ensure that any points between two encoded potentials in latent space can also
be decoded to a meaningful, scaled potential satisfying the original feature space distribution.
Therefore, we do not work with autoencoders as described above, but with so-called variational
autoencoders (VAEs), see Ref.~\cite{kingma2013auto}. VAEs introduce the concept of randomness to
the latent space. The central idea behind VAEs is that encoders do not directly map to a point but
rather a multivariate random distribution, in this case a Gaussian distribution
$\mathcal{N}(\bm{\mu}, \bm{\sigma})$, with distinct mean $\bm{\mu}\in\mathbb{R}^{L}$ and standard
deviation $\bm{\sigma}\in\mathbb{R}^{L}$ in latent space. From a technical perspective, the encoder
can be understood as a map $\mathbb{R}^F\to\mathbb{R}^{2L}$ that projects a vector
$\widetilde{\bm{U}}\in\mathbb{R}^F$ from feature space a tuple $(\bm{\mu}, \log(\bm{\sigma}))
\in\mathbb{R}^L\times\mathbb{R}^L$ with an elementwise logarithm.
By construction, the VAE learns to associate adjacent points in feature space with adjacent
points in latent space. 
Before entering the decoding pipeline, we need to keep in mind that the decoder is still a function
$\mathcal{D}:\mathbb{R}^L\to \mathbb{R}^F$, requiring an $L$-dimensional input. Therefore, we first
need to draw a single sample $\bm{l}\in\mathbb{R}^L$ from the above distribution 
$\mathcal{N}(\bm{\mu},\bm{\sigma})$, which then serves as the actual input to the decoder.
However, doing so naively would spoil the differentiability of the VAE and subsequent gradient-descent
based training. An alternative procedure that conserves differentiability and that comes at play here
is referred to as the reparametrization trick, see Ref.~\cite{kingma2013auto}: Let $\bm{\epsilon}\in
\mathbb{R}^{L}$ be a vector drawn from the multivariate standard distribution $\mathcal{N}(\bm{0},
\sum_i^L\bm{e}_i)$ in latent space. Then $\bm{l}=\bm{\mu} + \bm{\epsilon} \odot \bm{\sigma}$ can be
treated as a sample of $\mathcal{N}(\bm{\mu}, \bm{\sigma})$. Since $\bm{\epsilon}$ only enters as a
parameter, this is a valid starting point for decoding while still being differentiable with respect
to the encoder parameters.

If the VAE is has already been trained, there is no need in keeping this randomization, which would
only increase the model variance. In this case, especially during validation, we, do not apply
the reparametrization trick. Instead, we directly extract the bare mean $\bm{l}=\bm{\mu}$ as the
latent vector while ignoring the standard deviation $\bm{\sigma}$.

\subsubsection{Convolutional VAE Architecture}
Since we need to faithfully cover the correct short- and long-range behavior of $3\times 10^4$
training potentials in ${F=10^3}$ dimensional feature space, using deep encoders and decoders
is highly suggested. We decide to use fully-convolutional architectures, meaning that
the operation of classical pooling layers is also handled by convolutional layers with
increased strides, see Ref.~\cite{DB15a}. 

The  encoder and decoder architecture is shown in Figs.~\ref{fig:cvae_architecture}a) and b),
respectively. As the given scaled potentials $\widetilde{\bm{U}}\in X$ are merely treated as
vectors, we need to employ one-dimensional convolutional layers in both networks. At the beginning
of the encoding pipeline, a scaled potential is processed by two convolutional layers over a
total amount of $16$ channels with kernel size~$3$ and stride~$3$ in a row. Here as well as within
the decoder, convolutional layers are non-linearily activated by the GELU activation function,
see Ref.~\cite{hendrycks2016gaussian}. The application of each of these convolutional layers
reduces the amount of features per channel by a factor of three. Stacking all $16$ channels leaves
us with a $1776$ component vector. Since we choose to work with the latent space dimension $L=3$,
this vector is then mapped to a vector in $\mathbb{R}^6$ via a fully-connected layer, whose first
(last) three components are interpreted as $\bm{\mu}$ ($\bm{\sigma}$). Applying the reparametrization
trick for VAEs then yields the corresponding encoded potential. 

The decoder architecture is mainly a mirror of the encoder architecture: The encoded potential
is again mapped to a $1776$ component vector via a fully-connected layer, which is then split
into $16$ channels with each containing $111$ features. Analogously to the convolutional layers of
the encoder, we subsequently apply two transposed convolutional layers increasing the amount of
features per channel by a factor of three, instead. However, the second transposed convolutional
layer maps all incoming features to a single channel, which yields a $999$ component vector.
Finally, the remaining step of the reconstruction pipeline is to map this vector back to feature
space using another fully-connected layer.


\subsubsection{VAE Training}
\label{sec:vae-training}
\begin{figure}[t]
\includegraphics[width=0.98\columnwidth]{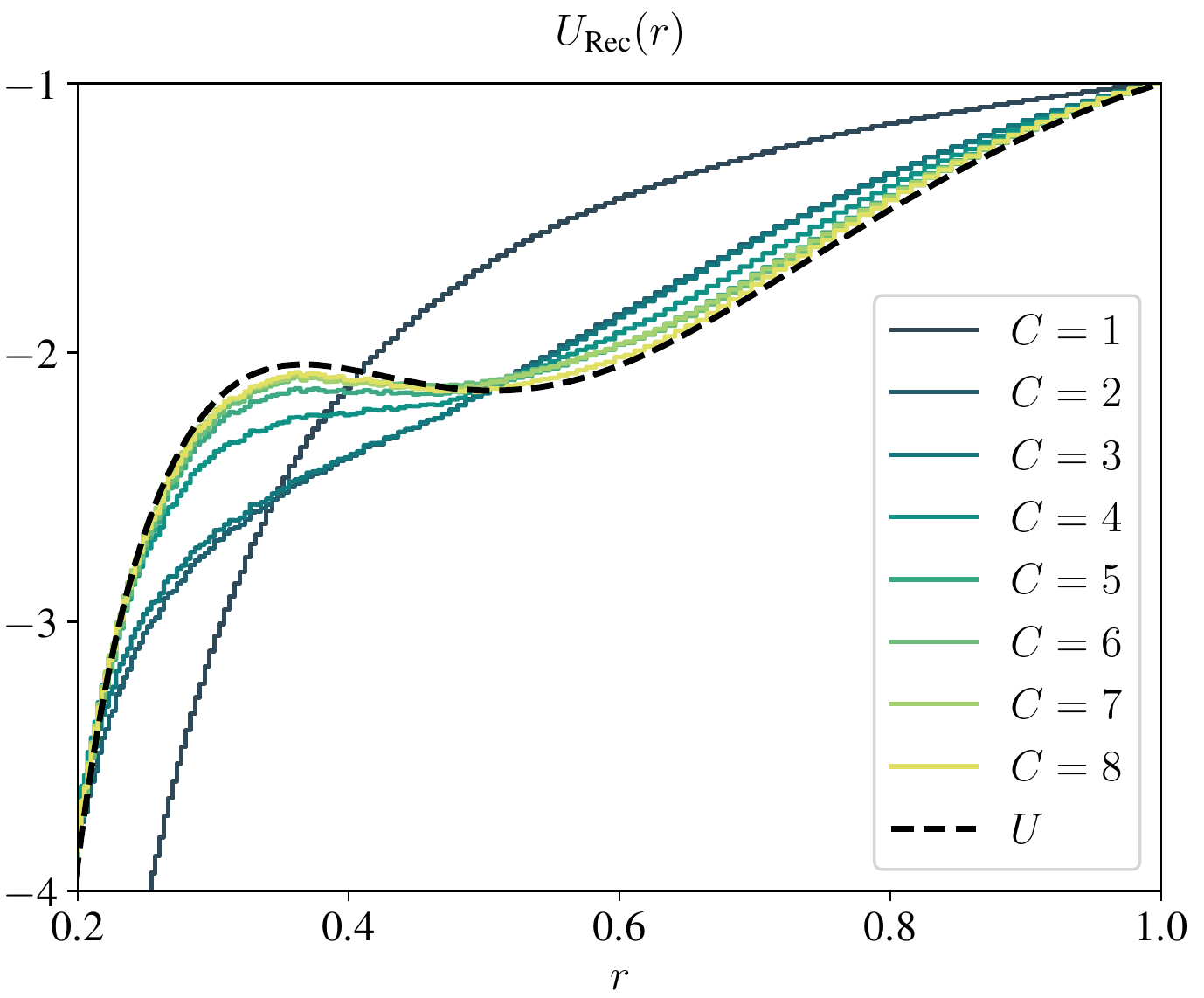}
\caption{Reconstruction of a random potential $U(r)$ (black dashed curve) for a varying size of the
autoencoder ensemble. A single VAE ($C1$) reproduces the overall trend only quite roughly, whereas
the maximal number of VAEs ($C=8$) provides the best approximation.}
\label{fig:cvae_contributions}
\end{figure}

While autoencoders are trained to minimize the reconstruction loss, there is an additional regulator
entering the objective function during the training of VAEs, known as the Kullback-Leibler-Divergence
$D_\text{KL}$ (KL-Divergence), see Ref.~\cite{10.2307/2236703}. In general, the KL-Diergence is a positive,
but asymmetric measure of how far two probability distributions deviate from each other. In the case of
VAEs, the KL-Divergence penalizes the encoder the more the distribution $\mathcal{E}(\widetilde{\bm{U}})
=\mathcal{N}(\bm{\mu}, \bm{\sigma})$ deviates from the multivariate standard distribution
$\mathcal{N}(\bm{0}, \sum_i^L\bm{e}_i)$ in latent space and is given by
\begin{equation}
D_\text{KL}(\widetilde{\bm{U}})=-\frac{1}{2}\sum\limits_{i=1}^L\left(1+2\log(\sigma_i) - \mu_i^2
- \exp(2\log(\sigma_i))\right),
\end{equation}
see Ref.~\cite{kingma2013auto}. The total loss that is minimized during the training of a VAE is the
sum of the reconstruction loss and the KL-Divergence. Ref.~\cite{higgins2016beta} introduces
an additional hyperparameter $\beta$ to tune the contribution of the KL-Divergence, which yields the loss
\begin{equation}
L(\widetilde{\bm{U}}) = L_\text{Rec}^{(\mathcal{A})}(\widetilde{\bm{U}}) + \beta D_\text{KL}(\widetilde{\bm{U}}).
\end{equation}
Working with $\beta\gg 1$ causes the VAE to priotize mapping any scaled potential $\widetilde{\bm{U}}\in X$
as closely as possible to a multivariate standard distribution in latent space. While this also provides
an encoded distribution $\mathcal{E}(X)$ that approximates a standard distribution quite well, this
severely reduces reconstruction accuracy. In contrast, the case $\beta\ll 1$ prioritizes accurate
reconstructions over having a symmetric, unimodal encoded distribution $\mathcal{E}(X)$ in latent
space. Ref.~\cite{10.2307/2236703} proposes a normalized factor $\beta_\text{norm}=\beta L/F$, which
is proportional to the quotient of latent and feature space dimensions. Although in our case
$L/F = 3\times 10^{-3}$, we work with an even smaller factor of $\beta=10^{-4}$. This yields
more accurate reconstructions while still producing sufficiently Gaussian encoded distributions,
as shown in Fig.~\ref{fig:cvae_contributions}.

The actual training pipeline is fairly straightforward. The VAE is trained on a standardized
data set containing the potentials $\widehat{\bm{U}}$ with components $\widehat{U}_i=(\widetilde{U}_i
-\mu^{(X)}_i)/\sigma^{(X)}_i$. Here, $\bm{\mu}^{(X)}$ and $\bm{\sigma}^{(X)}$ denote the elementwise mean
and, respectively, standard deviation of all $\widetilde{\bm{U}}\in X$. Afterwards, we train the
VAE over $N_E=10$ epochs using batch learning with batch size $30$ and the Adam optimizer,
see Ref.~\cite{kingma2014adam}. In order to obtain more stable results, we apply an exponentially
decaying learning rate schedule $\eta_i=10^{-3}/2^{i-1}$. Finally, we are not quite satisfied with
the resulting reconstruction loss of $3.35\times 10^{-2}$ on the test set $Y$, yet. This is
because the VAE only manages to reproduce the superficial behavior of input potentials, as shown in
Fig.~\ref{fig:cvae_contributions}. Using gradient boosting, see Sec.~\ref{sec:boosted-vae}, it is possible
to drastically reduce the reconstruction loss.

\subsubsection{Boosted VAE Ensemble}
\label{sec:boosted-vae}
Due to the considerably low latent space dimension $L=3$ the encoder shown in
Fig.~\ref{fig:cvae_architecture}a) severely compresses incoming information. The whole VAE architecture
can, thus, be understood as an extremely narrow information bottleneck. Consequently, the
decoder's capacity to reconstruct not only the rough behavior of the input potential, but also
finer oscillations, especially in the long-range regime, is fairly limited. This problem, however,
can be remedied by a boosting-based approach. Boosting techniques combine several of such weak
learners to a single strong learner, see Refs.~\cite{10.1023/A:1022648800760,6789696}. For the
given regression task of reconstructing scaled potentials $\widetilde{\bm{U}}\in X$, we apply
gradient boosting, see Ref.~\cite{10.2307/2699986}: We sequentially train $C$ autoencoders
$\mathcal{A}_1,\ldots,\mathcal{A}_C$ such that the $i^\text{th}$ member $\mathcal{A}_i$ reconstructs
the residuals of the previous VAE $\mathcal{A}_{i-1}$: 
\vfill
\noindent
\begin{minipage}{\linewidth}
\begin{algorithm}[H]
  \caption{Boosted VAE Ensemble - Training}
  \label{alg:boosted-vae-ensemble-train}
   \begin{algorithmic}[1]
	\Require{Training set $X\subseteq \mathbb{R}^F$, test set $Y\subseteq \mathbb{R}^F$} 
	\State $\mathcal{L}\leftarrow$ \textbf{empty list} \Comment{list of test reconstruction losses}
  \State $T\leftarrow (X, Y)$ \Comment{training and test sets}
	\State $C\leftarrow 0$ \Comment{number of VAEs in ensemble}
	\While{$\mathcal{L}$ does not converge}
	\State $C\leftarrow C+1$
	\State \textbf{initialize} VAE $\mathcal{A}_C$
	\State \textbf{train} $\mathcal{A}_C$ \textbf{on} $T_1$
	\State \textbf{append} $\frac{1}{|T_2|}\sum_{\widetilde{\bm{U}}\in T_2}\mathcal{L}_\text{Rec}^{(\mathcal{A}_C)}(\widetilde{\bm{U}})$ \textbf{to} $\mathcal{L}$ 
	\State $T\leftarrow (T_1-\mathcal{A}_C(T_1), T_2-\mathcal{A}_C(T_2))$
	\EndWhile
	\State \textbf{return} $\mathcal{A}_1,\ldots,\mathcal{A}_C$ \Comment{boosted VAE ensemble}
   \end{algorithmic}
\end{algorithm}
\end{minipage}\newpage

\noindent
Hence, each of the subsequent VAEs acts as a correction to its predecessor. This yields a
hierarchical sequence $\mathcal{A}$ of VAEs acting on scaled potentials as follows:
\begin{minipage}{\linewidth}
\begin{algorithm}[H]
  \caption{Boosted VAE Ensemble - Reconstruction}
  \label{alg:boosted-vae-ensemble-rec}
   \begin{algorithmic}[1]
	\Require{Pretrained VAEs $\mathcal{A}_1,\ldots,\mathcal{A}_C$} 
	\Ensure{Scaled potential $\widetilde{\bm{U}}$}
	\Function{$\mathcal{A}$}{$\widetilde{\bm{U}}$}
	\State $x\leftarrow\widetilde{\bm{U}}$
	\State $r\leftarrow 0$
	\For{$i\in\{1,\ldots,C\}$}
        \State $y\leftarrow\mathcal{A}_i(x)$
				\State $r\leftarrow r+y$
				\State $x\leftarrow x-y$
  \EndFor
	\State \textbf{return} $r$ \Comment{ensemble-reconstruction of $\widetilde{\bm{U}}$}
	\EndFunction
   \end{algorithmic}
\end{algorithm}
\end{minipage}\\

After having trained the eighth VAE $\mathcal{A}_8$, we do not observe any accuracy improvement when
adding further VAEs to the ensemble. Due to $C=8$, the effective latent space dimension is $C\times L = 24$.
As a result of gradient boosting we could reduce the reconstruction loss to $1.18\times 10^{-3}$,
which is more than one order of magnitude less than for the case of a single VAE.
Fig.~\ref{fig:cvae_contributions} shows how the reconstruction of some example potential
$\widetilde{\bm{U}}\in X$ improves for increasing $C$. While the true behavior of $\widetilde{\bm{U}}$
can only be guessed from the reconstruction via a single VAE $(C=1)$, the ensemble with the
maximum number of VAEs $(C=8)$ reliably reproduces oscillations in the long-range regime.

It is important to note that the individual VAE's latent spaces are independent of each other.
Hence, there is no way to express an ensemble-encoded potential by a single point in latent space,
but rather by a sequence ${\zeta = (\bm{\zeta}_1,\ldots,\bm{\zeta}_{C}) \in\mathbb{R}^{C\times L}}$
of $C$ points in latent space that we refer to as a latent curve. Here, the $i^\text{th}$ point
$\bm{\zeta}_i$ is the encoded contribution of $\mathcal{A}_i$ to the reconstruction of some given
potential $\widetilde{\bm{U}}$. As a consequence, ensemble-encoding satisfies a similar
formulation as Algorithm~\ref{alg:boosted-vae-ensemble-rec},
\begin{minipage}{\linewidth}
\begin{algorithm}[H]
  \caption{Boosted VAE Ensemble - Encoding}
  \label{alg:boosted-vae-ensemble-enc}
   \begin{algorithmic}[1]
	\Require{Pretrained VAEs $\mathcal{A}_1,\ldots,\mathcal{A}_{C}$ with $\mathcal{A}_i=\mathcal{D}_i\circ\mathcal{E}_i$} 
	\Ensure{Scaled potential $\widetilde{\bm{U}}$}
	\Function{$\mathcal{E}$}{$\widetilde{\bm{U}}$}
	\State $x\leftarrow\widetilde{\bm{U}}$
	\State $\zeta\leftarrow$ \textbf{empty list}
	\For{$i\in\{1,\ldots,C\}$}
				\State \textbf{append} $\mathcal{E}_i(x)$ \textbf{to} $\zeta$
        \State $y\leftarrow\mathcal{A}_i(x)$
				\State $x\leftarrow x-y$
  \EndFor
	\State \textbf{return} $\zeta$ \Comment{ensemble-encoding of $\widetilde{\bm{U}}$}
	\EndFunction
   \end{algorithmic}
\end{algorithm}
\end{minipage}
\begin{figure}[t]
\includegraphics[width=\columnwidth]{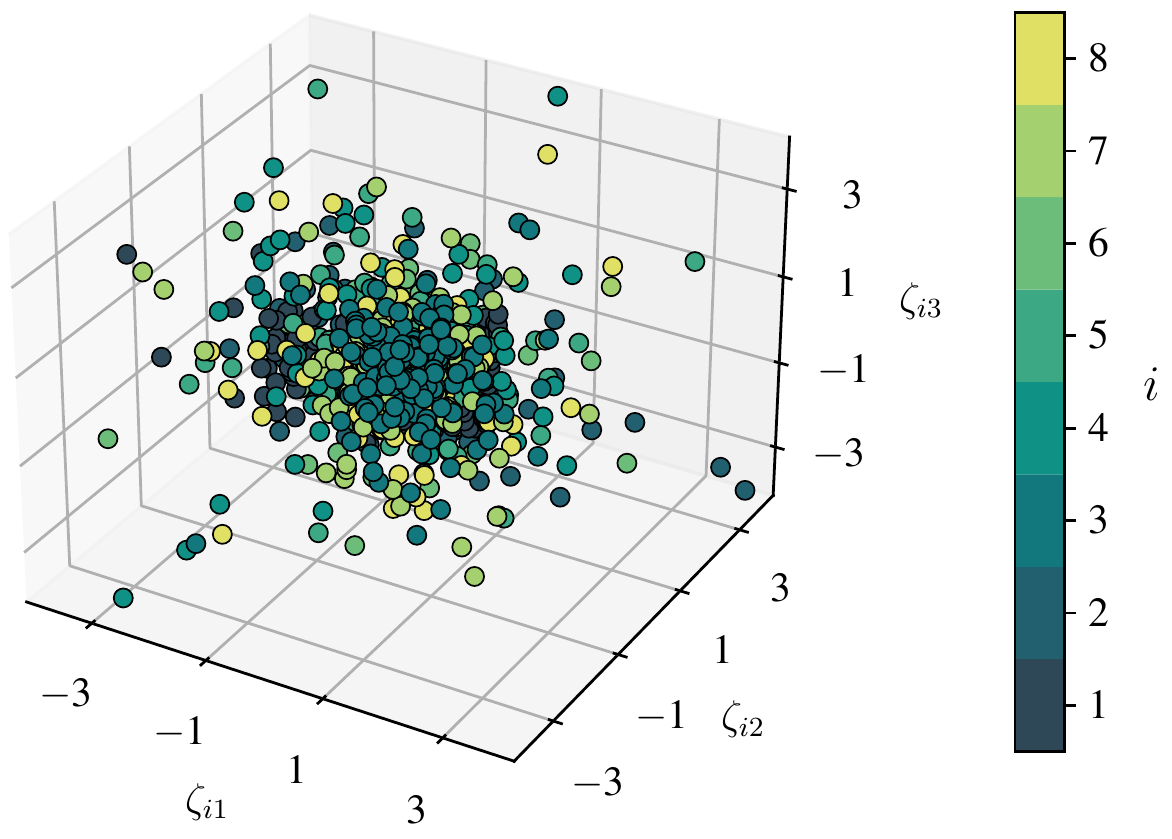}
\caption{Encoded contributions $\bm{\zeta}_i$ of all eight VAEs for $200$ random test potentials
in the respective three dimensional latent space spanned by the axes $\zeta_{i1}$, $\zeta_{i2}$,
$\zeta_{i3}$ with $i=1,\ldots,8$. All eight distributions are centered around the origin and
have standard deviations taking values $0.8-0.9$ and, therefore, roughly resemble standard distributions.}
\label{fig:latent_cloud}
\end{figure}

\noindent
Similar to a single encoder,  we can relate $X$ to a set $\mathcal{E}(X)\subseteq \mathbb{R}^{C\times L}$
of latent curves. 
When working with an ensemble of $C$ VAEs, the effective latent space dimension is, therefore, $C\times L$.
For our exploratory approach we are particularly interested in directly generating synthetic potentials
from the latent space. Given the latent curve ${\zeta = (\bm{\zeta}_1,\ldots,\bm{\zeta}_{C}) \in
\mathbb{R}^{C\times L}}$, the decoded potential is the sum of all decoded contributions,
\begin{equation}
\mathcal{D}(\zeta)=\sum\limits_{i=1}^{C} \mathcal{D}_i(\bm{\zeta}_i). \label{eq:ensemble-decode}
\end{equation}
\begin{figure}[t]
\includegraphics[width=0.95\columnwidth]{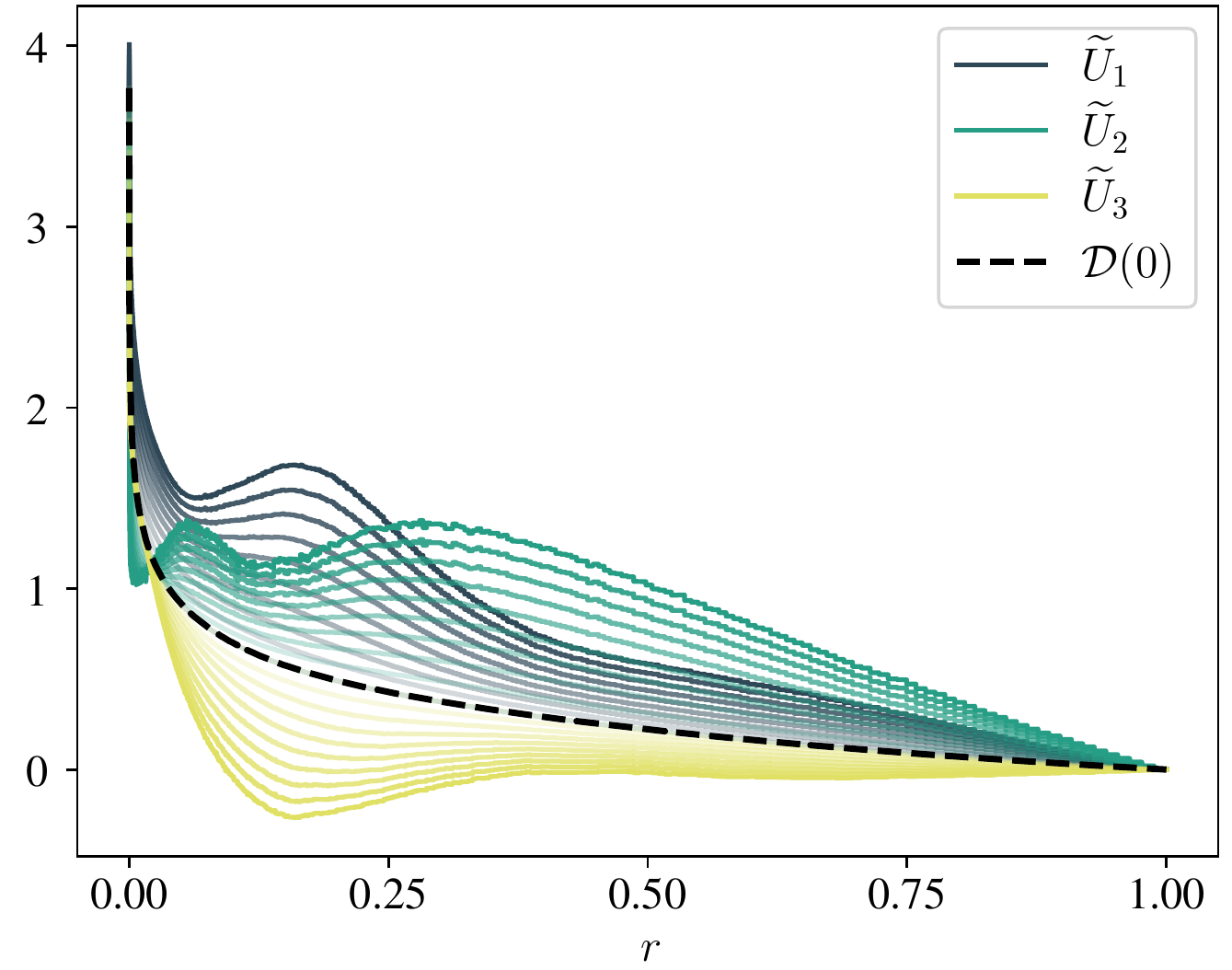}
\caption{Three different families of synthetically generated, scaled two-body potentials. At first,
three random latent curves are decoded, which yields the scaled potentials $\widetilde{U}_1$,
$\widetilde{U}_2$ and $\widetilde{U}_3$. In ten equidistant steps, these three latent curves
successively approach the origin of latent space. Decoding these provide the more opaque potentials.
Finally, the dashed black curve displays the decoded latent space origin, $\mathcal{D}(0)$, which can
be understood as the average potential of the underlying training set.}
\label{fig:synthetic_potentials}
\end{figure}
\begin{figure}[t]
\includegraphics[width=0.95\columnwidth]{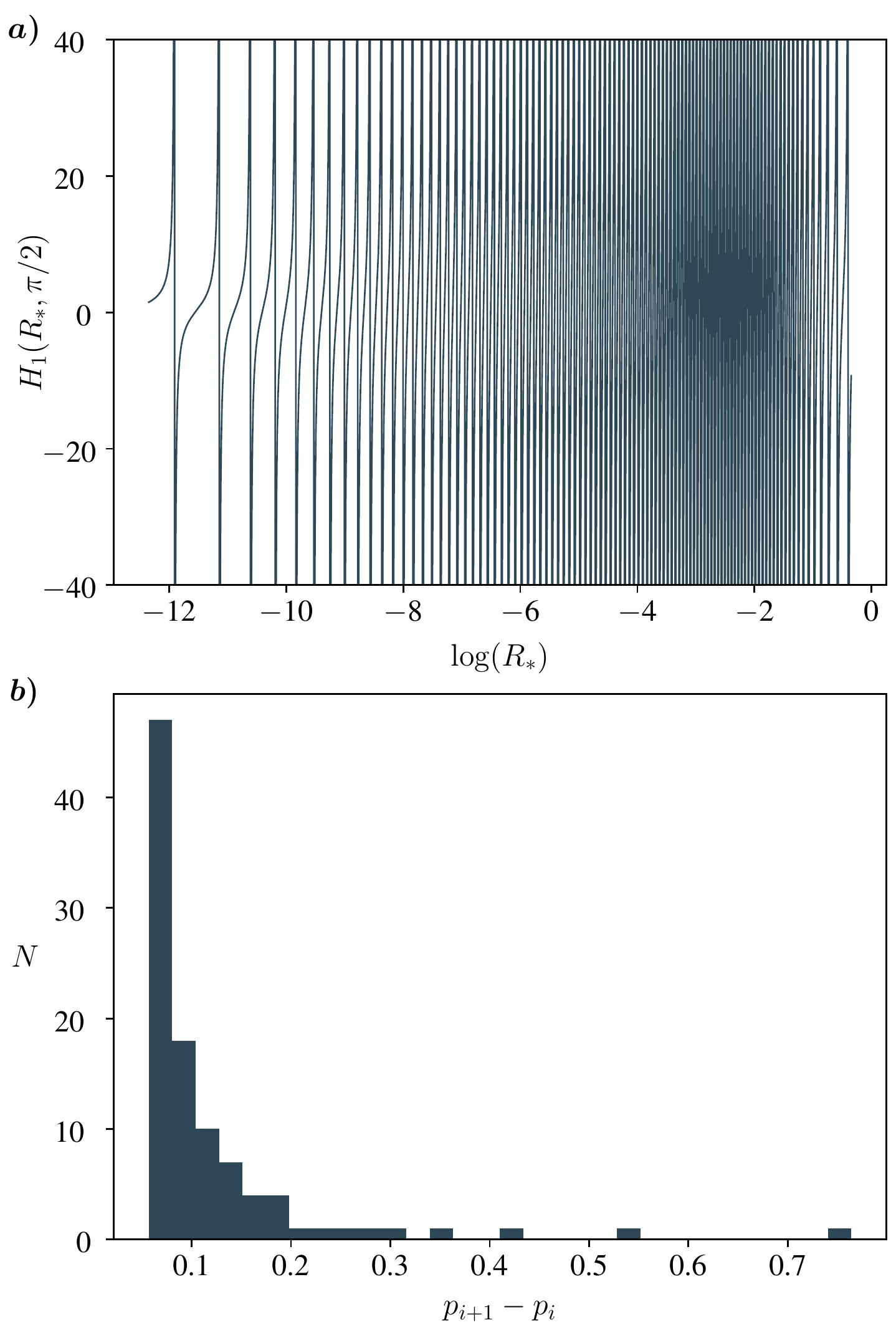}
\caption{a) displays the coupling constant $H_1(R_*,\pi/2)$ obtained from delta-shell renormalization
for a synthetic potential at hyperangle $\alpha=\pi/2$. The poles are clearly not equidistant, which
violates discrete scaling symmetry and indicates that the given potential does not give rise to a limit
cycle.
b) that shows how the distances $\Delta x$ between adjacent poles are distributed. The width of
this distribution can be used to measure the deviation from a limit cycle.}
\label{fig:lc_error}
\end{figure}
\noindent
Knowing the underlying latent curve distribution $\mathcal{E}(X)$ is essential for generating synthetic
potentials using Eq.~\eqref{eq:ensemble-decode}. Fig.~\ref{fig:latent_cloud} displays the distribution
of $\bm{\zeta}_i$ for each autoencoder $\mathcal{A}_i$ and a sample of $200$ latent curves
${\zeta=(\bm{\zeta}_1,\ldots,\bm{\zeta}_8)\in \mathcal{E}(X)}$. The shown distributions strongly
resemble each other. When considering the whole set $\mathcal{E}(X)$ we find each distribution being
centered closely to the origin and having standard deviations ranging between $0.8$ and $0.9$, depending
on $i$ and the considered axis in latent space. We will, therefore, approximate the latent curve distribution
by a multivariate standard distribution. Fig.~\ref{fig:synthetic_potentials} shows three different families
of synthetic potentials. The three main curves $\widetilde{U}_1(r)$, $\widetilde{U}_2(r)$ and
$\widetilde{U}_3(r)$ have been generated from a standard distribution in $\mathbb{R}^{C\times L}$.
The dashed curve is the decoded origin of $\mathbb{R}^{C\times L}$ and can be understood as an average
representant of $X$.

\subsection{Downstream task}
\label{sex:downstream-task}
\subsubsection{Limit-cycle-loss}
\label{sec:limit-cycle-loss}
The boosted VAE ensemble from Sec.~\ref{sec:boosted-vae} provides a dimensionality reduction which
greatly simplifies the search for LC potentials in latent space. Having specified the latent space
as the actual search space, this section is dedicated to the motivation of the limit-cycle-loss (LCLoss):
For a latent curve $\zeta\in\mathbb{R}^{C\times L}$ the LCLoss $\mathcal{L}_\text{LC}$ determines how
much the coupling constant $H_n(R_*,\alpha)$ of the potential $\mathcal{D}(\zeta)$ deviates from
a log-periodic behavior. Therefore, the LCLoss has to be understood as a function
$\mathcal{L}_\text{LC}:\mathbb{R}^{C\times L}\to \mathbb{R}_0^+$. If and only if $\mathcal{L}_\text{LC}(\bm{\zeta})
=0$, the potential $\mathcal{D}(\zeta)$ has to be LC. For simplicity we choose a fixed hyperangle
$\alpha=\pi/2$ and node index $n=1$.
Figs.~\ref{fig:lc_error}a) and b) demonstrate how the LCLoss for some $\zeta\in\mathbb{R}^{C\times L}$
is computed: The idea is to find all poles $p_i$ of $H_n(R_*,\alpha)$ in the interval ${\log(R_*)
\in[-12-\log(\sqrt{2}\sin(\alpha)), -\log(\sqrt{2}\sin(\alpha))]}$ (Fig.~\ref{fig:lc_error}a).
Considering the pairwise distances $p_{i+1}-p_i$ between adjacent log-poles yields some distribution
$\Lambda$, see Fig.~\ref{fig:lc_error}b). If $\mathcal{D}(\zeta)$ is LC, then all poles are
equidistant, which causes all pairwise distances to be identical and the standard deviation
$\sigma(\Lambda)$ to vanish. In this case, $\log(\lambda)=\mu(\Lambda)$ is the log-periodicity of
$H_n(R_*,\alpha)$, which is related to the preferred scaling factor $\lambda$ of that limit cycle.
In general, larger LCLosses indicate larger deviations from a log-periodic behavior. This makes the quotient
\begin{equation}
\mathcal{L}_\text{LC}(\zeta) = \frac{\sigma(\Lambda)}{\mu(\Lambda)} \label{eq:lcloss}
\end{equation}
a promising candidate for the required measure. However, there is one difficulty that still needs to
be addressed. Namely, if $H_n(R_*,\alpha)$ is not singular or if it gives rise to a small number
$P\sim \mathcal{O}(1)$ of poles, then the LCLoss defined as in Eq.~\eqref{eq:lcloss} is no
reliable measure for LC-ness. The same problem holds if the density of all poles is of the same
order as the hyperradial resolution, that is $\mu(\Lambda)\approx 1.2\times 10^{-4}$, which corresponds
to a number of $P\approx 10^5$ poles. Therefore, we require 
\begin{equation}
0\ll P \ll 10^5 \label{eq:p-inequality}
\end{equation}
to avoid an ill-defined LCLoss.
Since all scaled potentials $\widetilde{\bm{U}}$ in $X$ and $Y$ have been generated to satisfy the
normalization $\widetilde{U}(1)=0$, that is $U(r)=-1$, there is one additional degree of freedom we
have ignored until now. In the following we introduce an additional displacement $s$ that is fed
to the rescaled potential as follows:
\begin{equation}
U(r)=-\mathrm{e}^{8(\widetilde{U}(r)+s)} \label{eq:scale-operation}
\end{equation}
This corresponds to scaling the original potentials by the factor $\mathrm{e}^{8s}$.
While the number of poles increases with larger $s$, the shape of the distribution $\Lambda$ turns
out to be invariant under this operation: An LC (non-LC) potential stays LC (non-LC), regardless of
the value of $s$. 
An appropriate choice of the displacement $s$ for some latent curve $\bm{\zeta}$ allows an
alternative normalization in which $H_n(R_*,\alpha)$ has only a fixed number $P=10^2$ of poles and
which is, therefore, more compatible with the above definition of the LCLoss.
The relation between latent curves $\zeta\in\mathbb{R}^{C\times L}$ and the correct displacements is
established via a supervisedly trained ensemble $\mathcal{S}$ of $C=8$ CNNs
$\mathcal{S}_1,\ldots,\mathcal{S}_C:\mathbb{R}^{C\times L}\to\mathbb{R}$. Each CNN shares the same
architecture as the decoders $\mathcal{D}_i$, see Fig.~\ref{fig:cvae_architecture}b). In fact,
the only difference lies within the output layer, which maps to a real number $s\in\mathbb{R}$
instead of some $\widetilde{\bm{U}}\in\mathbb{R}^F$. The loss function to be minimized during the 
supervised training of the $\mathcal{S}_i$ is simply the L1Loss between targets and predictions, as defined in Eq.~\eqref{eq:l1loss}.

The training and test sets, $X_\mathcal{S}$ and $Y_\mathcal{S}$, we use to train the ensemble
$\mathcal{S}$ contain $|X_\mathcal{S}|=3\times 10^3$ and, respectively, $|Y_\mathcal{S}|=3\times 10^2$
pairs $(\bm{\zeta}, s)$. For each latent curve ${\zeta\in\mathbb{R}^{C\times L}}$ the corresponding
displacement $s$ is found via a grid search among the $150$ equidistant displacements $s_i=-5+(i-1)/149$.
As the number of poles strictly increases monotonically in terms of the displacement $s$, there
is a unique solution for each latent curve.

$\mathcal{S}$ undergoes a similar boosting procedure like the VAE ensemble $\mathcal{A}$,
see Algorithm~\ref{alg:boosted-vae-ensemble-train}. Of course, it is important to realize that
during the $i^\text{th}$ iteration we only adapt the targets of $T_1$ and $T_2$ to the previous
residuals $s-\sum_{j=1}^{i-1}\mathcal{S}_j(\zeta)$, while the given latent curves remain unchanged.
Then the resulting ensemble-prediction on the displacement $s$ is the sum
\begin{equation}
\mathcal{S}(\zeta) = \sum\limits_{i=1}^C \mathcal{S}_i(\zeta)
\end{equation}
For the individual CNNs $\mathcal{S}_i$ a similar training pipeline as in Sec.~\ref{sec:vae-training}
proves to be useful. We only introduce minor changes like an additional weight decay of $10^{-3}$
and noise injection with the standard deviation $0.025$ for further regularization as well as a
modified learning rate schedule with $\eta_i=10^{-3}\cdot 0.95^{i-1}$. To verify that the ensemble
$\mathcal{S}$ meets the requirements, we generate $10^3$ synthetic potentials $\widetilde{\bm{U}}
=\mathcal{D}(\zeta)\in\mathbb{R}^F$ from a standard distribution in $\mathbb{R}^{C\times L}$.
After rescaling each potential to
\begin{equation}
U(r)=-\mathrm{e}^{8(\widetilde{U}(r)+\mathcal{S}(\zeta))} \label{eq:rescaled-s-ensemble}
\end{equation}
we count all poles of the coupling constant $H_n(R_*,\alpha)$ for $\alpha=\pi/2$ and $n=1$. The resulting
distribution of pole numbers $P$ is approximately Gaussian, having the mean $\mu=101.8$ and
standard deviation $\sigma=21.2$. Indeed, the standard deviation is not that small,
compared to the mean. Nevertheless, using the rescaled potentials from Eq.~\eqref{eq:rescaled-s-ensemble}, 
the ensemble $\mathcal{S}$ still suffices for a sane definition of the LCLoss in Eq.~\eqref{eq:lcloss}.
This is because virtually each synthetic potential that is generated during the downstream task
exhibits the right amount $P$ of poles in its coupling constant to satisfy Eq.~\eqref{eq:p-inequality}.

\subsubsection{Genetic algorithm for a latent curve population}
\label{sec:ga}
Since the LCLoss depends on the exact pole positions of the coupling constant $H_n(R_*, \alpha)$,
its loss surface is neither smooth, nor convex, which renders optimization approaches based on
gradient descent fairly difficult. Instead, we decide to search for LC potentials using an
elitist genetic algorithm (GA), which is motivated by Goldbergs's Simple Genetic Algorithm,
see Ref.~\cite{goldberg1989genetic}. 

The very first step of each GA is to initialize a population. Here, we draw $100$ latent curves
(the individuals) from a standard distribution in $\mathbb{R}^{C\times L}$, which all together form
the initial population. Then we compute the LCLoss for the corresponding rescaled potentials
from Eq.~\eqref{eq:rescaled-s-ensemble}. The main part of the GA is organized in generations,
where each generation consists of a sequentially executed selection, crossover, mutation and
removal phase. Using standard genetic operators we aim at evolving the population towards fitter
individuals, that is potentials with lower LCLosses, within $100$ generations.

At the beginning of each generation, we identify the fittest individual $\zeta_\text{fittest}$,
which is the latent curve with the lowest LCLoss. $\zeta_\text{fittest}$ is reserved for crossover and
cannot be eliminated during removal phase.
Within the selection phase, we carry out ten tournament selections with tournament size $20$ among
all individuals but the fittest one. Once an individual has won a tournament due to having the
lowest LCLoss compared to the other $19$ competitors, it is reserved for crossover and cannot
participate in the following tournaments.
During the crossover phase, the fittest individual $\zeta_\text{fittest}$ mates with each of the
ten individuals $\zeta_i$ with $i=1,\ldots,10$ that won a tournament during selection phase.
Each couple $(\zeta_\text{fittest}, \zeta_i)$ generates two children $\zeta$ via a heuristic crossover,
\begin{equation}
\zeta = \zeta_\text{fittest} + r (\zeta_\text{fittest} - \zeta_i)
\end{equation}
with $r$ drawn from the uniform distribution $\mathcal{U}([0,1])$. At this step, the population
consists of $120$ individuals. Due to being extremely elitist, this GA is prone to converge against
local minima.  In order to counteract this issue, we enhance the genetic diversity, which can be
understood as width of the population in $\mathbb{R}^{C\times L}$, during mutation phase. Gaussian mutation
appears to suffice these diversification requirements and is applied to the offspring provided by
the crossover phase as follows:
\begin{equation}
\zeta_\text{mutated} = f\odot\zeta,
\end{equation}
where $f\in\mathbb{R}^{C\times L}$ is drawn from the multivariate normal distribution $\mathcal{N}(1,\sigma_g)$
with $g$ denoting the number of past generations. We apply an exponentially decaying mutation parameter
$\sigma_g=0.5 \cdot 0.98^{g}$, which causes mutation to become less relevant compared to crossover
towards later generations. In doing so, we assume that the GA finds well performing individuals
during early generations and afterwards only needs to perform a crossover-based fine-tuning.
At the end of the generation, we carry out $20$ tournament selections with tournament size $20$.
However, in contrast to the tournaments during selection phase, we remove the individual with the
highest LCLoss from the population, such that we are left with $100$ individuals at the end of
each generation.

The GA as described above is not performed once, but in parallel for $50$ populations that evolve
independently from each other, instead. After fulfilling the stopping criterion of reaching the
$101^\text{st}$ generation, we extract the fittest individuals from each population.
\begin{figure}[t]
\includegraphics[width=0.95\columnwidth]{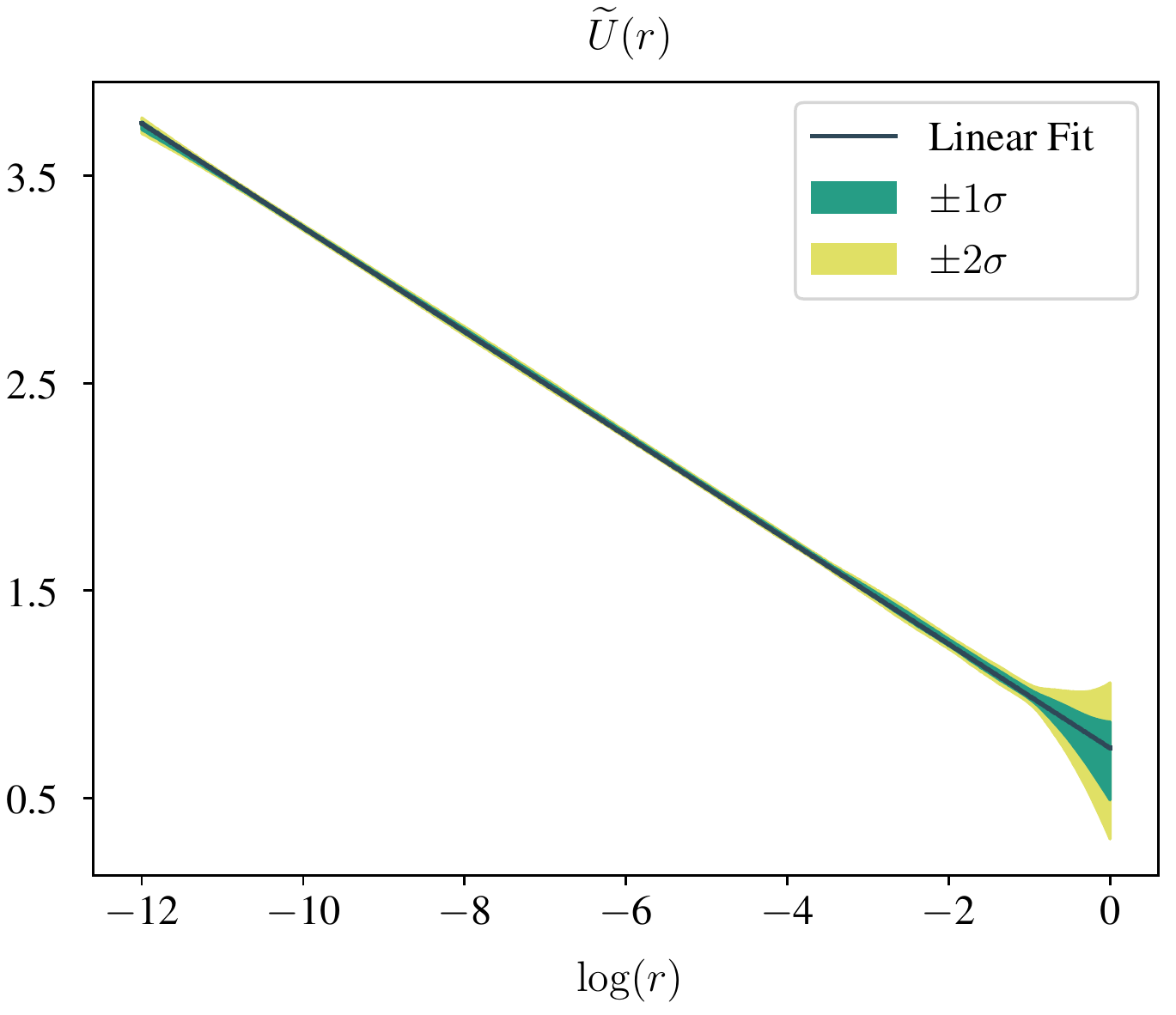}
\caption{Linear model $\widetilde{U}(r)=a\log(r)+s$ with fit parameters from Eq.~\eqref{eq:linear-fit}
(black curve) as well as $1\sigma$ and $2\sigma$ level for the distribution the fittest individuals
from all $50$ GAs. The fittest individuals are very similar to each other, especially
in the short-range regime, and behave log-linearily.}
\label{fig:fittest-from-ga}
\end{figure}

\section{Results}
\label{sec:results}
Whether the search for LC potentials is successful depends primarily on the results of the downstream
task. While the boosted VAE and CNN ensembles $\mathcal{A}$ and $\mathcal{S}$ merely provide the
theoretical framework to reliably compute LCLosses in a low-dimensional representation, the actual
search is carried out by the GAs in Sec.~\ref{sec:ga}. For this reason, the following analysis is
heavily based on the fittest individuals that are produced by the latter.
\subsection{Fittest individuals from genetic algorithm}
\label{sec:fittest-individuals}
The fact that we have applied the above GA not to one, but to fifty independent populations in parallel
allows to estimate a distribution $Z$ of latent curves in $\mathbb{R}^{C\times L}$, based on the corresponding
fittest individuals. In order to save computational resources, we want to avoid carrying out the same 
GA explicitly to further populations and
subsequently extracting the fittest individuals in each case. Instead, we directly draw new latent curves 
$\zeta$ from the distribution $Z$, which have expectedly low and can compete with the fifty extracted
latent curves from Sec.~\ref{sec:ga}.
Fig.~\ref{fig:fittest-from-ga} displays the $1\sigma$ and $2\sigma$ levels of that distribution.
It is remarkable how similar the corresponding scaled potentials $\widetilde{\bm{U}}=\mathcal{D}(\zeta)$
are to each other, especially in the short-range regime, all $\widetilde{U}(r)$ behave notably linear
in terms of $\log(r)$. 

The mentioned similarity implies that the loss surface of the LCLoss does not exhibit several distinct
local minima that perform equally well, but one global minimum each GA draws its offspring towards,
instead. Our search for LC potentials, therefore, appears to yield a unique solution. The fittest of
all $50$ individuals is the latent curve with the lowest LCLoss of among all GAs. Its considerably
low LCLoss of $\mathcal{L}_\text{LC}=1.06\times 10^{-2}$ indicates the associated scaled potential to
be close to this unique LC potential.

The variance in the long-range regime is slightly increased, which can be traced back to the
behavior of the VAE ensemble and the nature of the LCLoss. Depending on the given latent curve,
the VAE ensemble may produce perturbations in the long-range sector. As long as the potential
ensures a coupling constant $H_n(R_*, \alpha)$ with almost equidistant log-poles in the short-range
sector, the loss gain due to these perturbations becomes negligibly small. Finally, in an actual
GA such an individual may still be superior compared to other individuals in the same population. 
To conclude, we attribute the non-linearity in the long-range sector to a lack of sensitivity of
the LCLoss to deviations from LC-ness in narrow $\log(R_*)$-intervalls. As a consequence,
we assume the demanded scaled potential to be a linear function in $\log(r)$.
Fitting a linear model $a\log(r)+s$ each of the $50$ individuals yields one distribution
per fit parameter. From these, we deduce
\begin{equation}
a = -0.251 \pm 0.003, \hspace{0.5cm} s = 0.742 \pm 0.029. \label{eq:linear-fit}
\end{equation}
The fit model with the parameters $a$ and $s$ from Eq.~\eqref{eq:linear-fit} is displayed as the
black curve in Fig.~\ref{fig:fittest-from-ga}. The corresponding LCLoss $\mathcal{L}_\text{LC}
=1.5\times 10^{-2}$ is even slightly larger than that of the previously mentioned fittest individual.
Nevertheless, it is still worth pursuing such linear scaled potentials as we will show in
Sec.~\ref{sec:different-hyperangles}. Note that rescaling a potential $\widetilde{U}(r)\propto
\log(r)$ leads to an $1/r^p$ potential,
\begin{equation}
\widetilde{U}(r)=a\log(r)+b \Longrightarrow U(r) \propto \mathrm{e}^{8a\log(r)} = 1/r^{p},
\end{equation}
which establishes a relation between the exponent $p$ and the slope $a$. Inserting the fit parameter
$a$ from Eq.~\eqref{eq:linear-fit} we find
\begin{equation}
p = -8a = 2.0086 \pm 0.0236. \label{eq:exponent}
\end{equation}

\subsection{$1/r^p$ potentials at different hyperangles}
\label{sec:different-hyperangles}
The exponent $p$ found in Sec.~\ref{sec:fittest-individuals} allows for some speculations regarding
the unique LC potential. The behavior ${\widetilde{U}(r)\propto \log(r)}$ suggests that it is of
type $1/r^p$. Since $p$ closely approximates the value $2$, the $1/r^2$ potential is a
promising candidate for the demanded potential. 
Until now, however, the LCLoss has only been computed for the specific hyperangle $\alpha=\pi/2$.
As a consequence, the results of the previous GAs do not provide sufficient evidence to
identify the $1/r^2$ potential as a general solution that minimizes the LCLoss independent of the
hyperangle.

Evaluating LCLosses of $1/r^p$ potentials over a two-dimensional grid of exponents and hyperangles
sheds light on this problem. Since cutoff hyperradii ${R_*=r_*/(\sqrt{2}\sin(\alpha))}$ are
inversely proportional to the hyperangle $\alpha$ while entering the arguments of the Bessel
functions in Eq.~\eqref{eq:lambdacoupling}, the number $P$ of poles inside the considered
interval ${[-12-\log(\sqrt{2}\sin(\alpha)), -\log(\sqrt{2}\sin(\alpha))]}$ diverges in the
limit $\alpha\to 0$. As the ensemble $\mathcal{S}$ does not provide an appropriate normalization
at other hyperangles than $\pi/2$, it cannot counteract the increase in $P$. Finally,
the pole density must not exceed the hyperangular resolution, which is why we need to
define a lower bound for hyperangles to be considered here. 

\begin{figure}[t]
\includegraphics[width=\columnwidth]{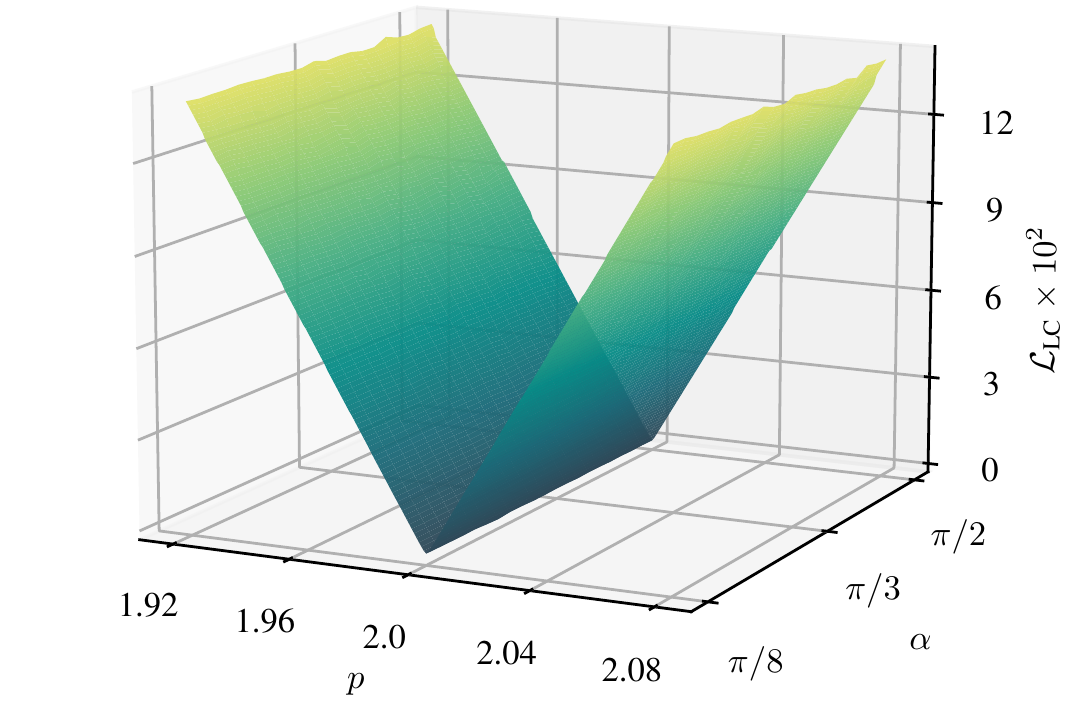}
\caption{LCLoss for $1/r^p$ potentials over the exponent $p$ and hyperangle $\alpha$. The lowest
losses are aligned at $p=2$ and take values between $\mathcal{L}_\text{LC} = 2.332\times 10^{-4}$
and $\mathcal{L}_\text{LC} = 9.389\times 10^{-4}$, suggesting the $1/r^2$ potential to
be the desired LC potential.}
\label{fig:exponents}
\end{figure}

Fig.~\ref{fig:exponents} displays the evaluated LCLosses over the rectangle $(p,\alpha)\in[1.92,2.08]
\times [\pi/8,\pi/2]$. We observe the shown loss surface to have a distinctive ravine for
the exponent $p=2$. Ranging from $\mathcal{L}_\text{LC} = 2.332\times 10^{-4}$ being the global
minimum to $\mathcal{L}_\text{LC} = 9.389\times 10^{-4}$, the losses along $p=2$ are all satisfactorily
low. The fact that the LCLoss does not vanish exactly is most likely a discretization artifact
and can be neglected for our purposes.

In summary, we determine the $1/r^2$ potential to be the desired LC potential as it minimizes the
LCLoss independent of the hyperangle to a value close to zero. Due to the observed similarity between
all fittest individuals from Sec.~\ref{sec:fittest-individuals}, we can exclude the existence of
other LC potentials. The topology of the loss surface shown in Fig.~\ref{fig:exponents} invites
to generalize our findings to smaller hyperangles as well. Fine-tuning the definition of the
LCLoss and making it suitable for smaller hyperangles, the $1/r^2$ potential seems also likely
to be the unique solution to our search for $\alpha\in[0,\pi/8]$. 

\subsection{Log-periodicity for the $1/r^2$ potential}
There are two mechanisms that control the number $P$ of poles the coupling constant $H_n(R_*,\alpha)$
has in the interval $[-12-\log(\sqrt{2}\sin(\alpha)), -\log(\sqrt{2}\sin(\alpha))]$.
When defining the LCLoss in Sec.~\ref{sec:limit-cycle-loss}, we already took advantage of the
first one, being the displacement operation introduced in Eq.~\eqref{eq:scale-operation}.
It corresponds to multiplying an additional factor $\exp(8s)$ to the rescaled potential and
increases the momenta $k_n(u)$ in Eq.~\eqref{eq:momenta-knu}. Entering the arguments of the Bessel
functions in Eq.~\eqref{eq:lambdacoupling}, this finally increases the pole density.
The second mechanism is fine-tuning the hyperangle $\alpha$, which has been briefly mentioned
in Sec.~\ref{sec:different-hyperangles}. Due to the inverse proportionality between cutoff
hyperradii $R_*$ and hyperangles $\alpha$, the pole density increases whenever $\alpha$ decreases.
Similar to the displacement operation, this happens due to the arguments of the Bessel
functions in Eq.~\eqref{eq:lambdacoupling} being proportional to~$R_*$.
\begin{figure}[t]
\includegraphics[width=\columnwidth]{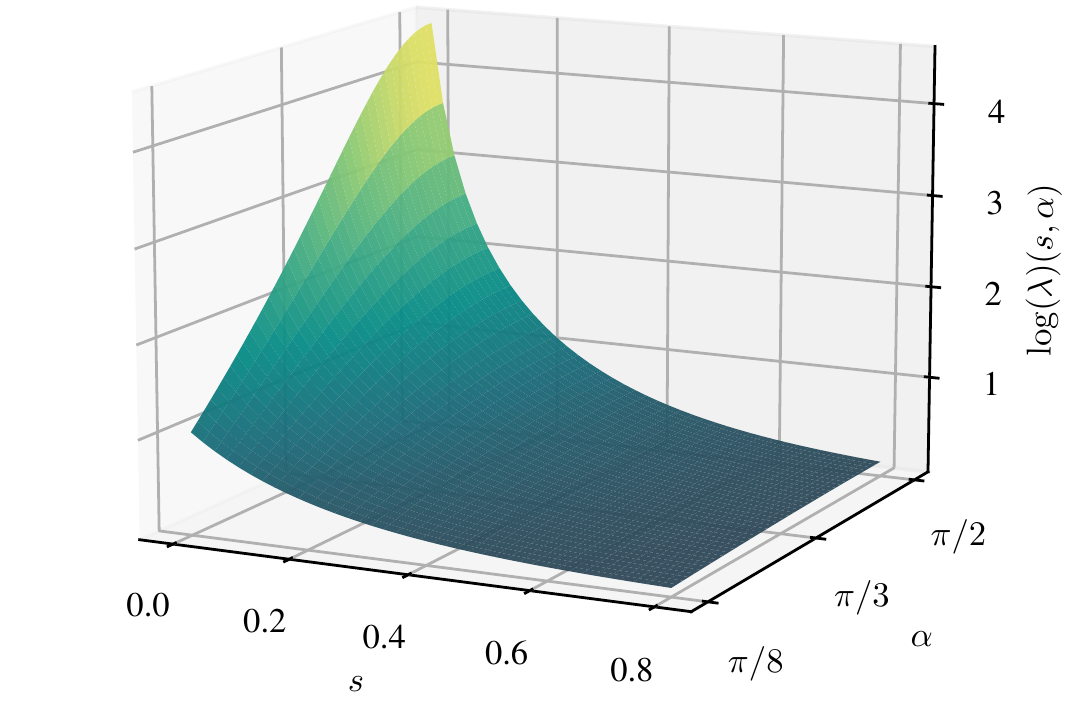}
\caption{Log-periodicity $\log(\lambda)(s,\alpha)$ of the coupling constant $H_1(R_*,\alpha)$
in terms of the displacement $s$ and hyperangle $\alpha$. Increasing $s$ leads to an approximately
exponential decay of $\log(\lambda)(s,\alpha)$ and, therefore, to a larger number $P$ of poles.
Vice-versa, large $\alpha$ cause $\log(\lambda)(s,\alpha)$ to increase, which reduces $P$, instead.}
\label{fig:logperiodicity}
\end{figure}
\noindent
The central result of our analysis is that the $1/r^2$ potential uniquely minimizes the LCLoss.
For each pair of displacement $s$ and hyperangle $\alpha$ we can, therefore, assign any $1/r^2$
potential to the log-periodicity $\log(\lambda)(s,\alpha)$ of its coupling constant $H_n(R_*,\alpha)$
with node index $n=1$. Fig.~\ref{fig:logperiodicity} shows how $\log(\lambda)(s,\alpha)$ depends on
$s$ and $\alpha$ and, thereby, visualizes both discussed mechanisms of steering
\begin{equation}
P(s,\alpha) \approx \left\lceil \frac{12}{\log(\lambda)(s,\alpha)} \right\rceil. \label{eq:p-estimate}
\end{equation} 
The behavior of $\log(\lambda)(s,\alpha)$ can be described by the model
\begin{equation}
\log(\lambda_\text{fit})(s, \alpha) = \exp\left(\exp(bs)\sum\limits_{i,j=0}^2 c_{ij} s^i\alpha^{2j+1}\right)-1,
\label{eq:nonlinear-fit}
\end{equation}
which suffices to cover both mechanisms. It cannot be expressed via a separation ansatz
${\log(\lambda_\text{fit})(s, \alpha)=f_1(s)f_2(\alpha)}$ due to containing several mixed terms.
Fitting this model to the data shown in Fig.~\ref{fig:logperiodicity} yields $b=-6.230\pm 0.027$
and the fit parameters $c_{ij}$ listed in Tab.~\ref{tab:c-coeffs}.
\begin{table}[h]
\centering
\begin{tabular}{|c|rrr|}
\hline
$c_{ij}$ & $j=0 \ \hspace{0.35cm} \ $ & $j=1 \ \hspace{0.35cm} \ $ & $j=2 \ \hspace{0.35cm} \ $ \\
\hline
$i=0$ & $1.709\pm 0.004$ & $-0.345\pm 0.005$ & $0.039\pm 0.002$ \\
$i=1$ & $5.775\pm 0.090$ & $-3.386\pm 0.121$ & $0.658\pm 0.039$ \\
$i=2$ & $10.422\pm 0.460$ & $1.954\pm 0.478$ & $-0.782\pm 0.155$ \\
\hline
\end{tabular}
\caption{Fit parameters $c_{ij}$ of the non-linear model $\log(\lambda_\text{fit})$ in
Eq.~\eqref{eq:nonlinear-fit} fitted to the data shown in Fig.~\ref{fig:logperiodicity}.}
\label{tab:c-coeffs}
\end{table}
\begin{figure}[t]
\includegraphics[width=0.96\columnwidth]{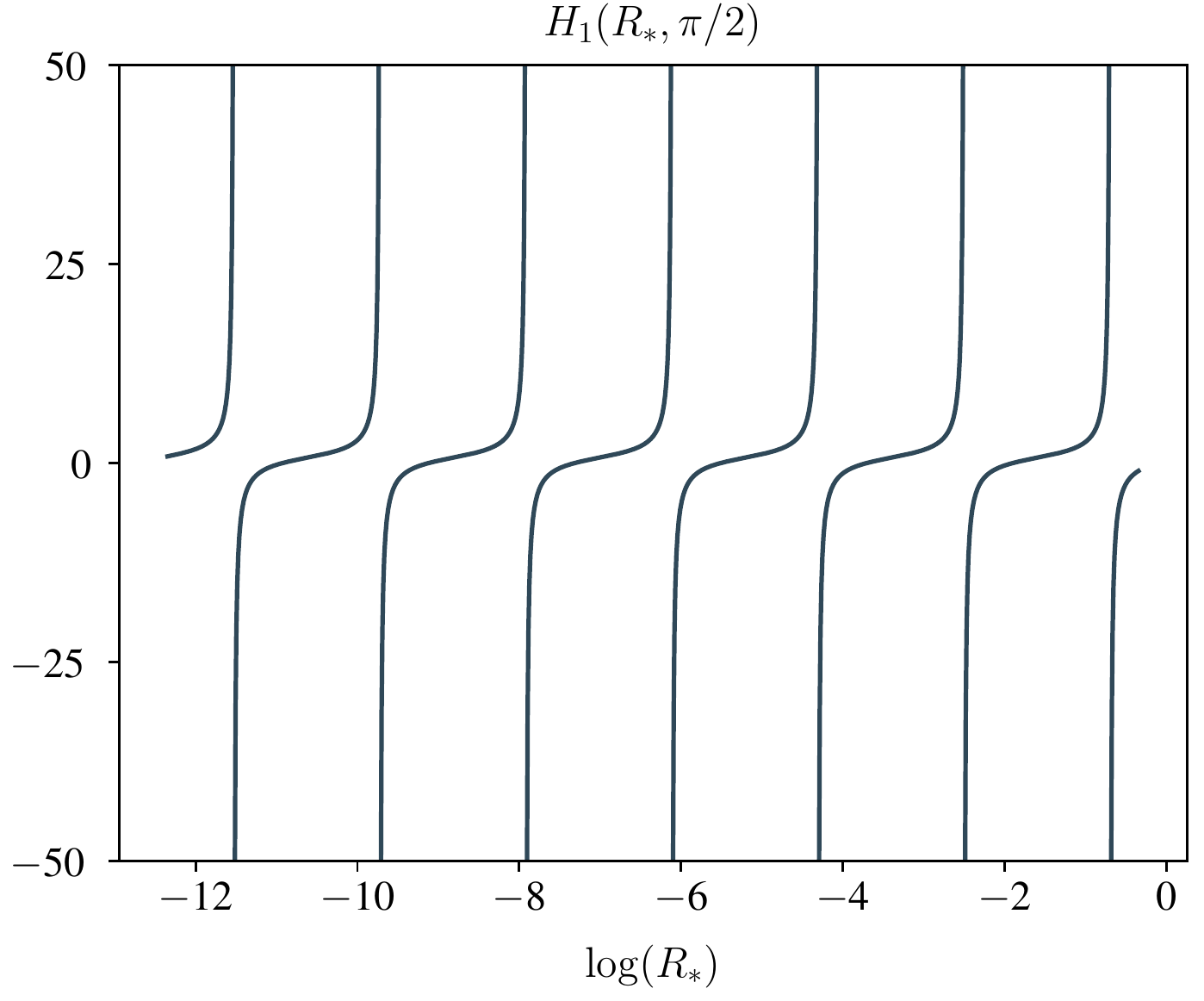}
\caption{Coupling constant $H_1(R_*,\pi/2)$ for the $1/r^2$ potential with displacement $s=1/8$
over the logarithmic cutoff hyperradius $\log(R_*)$, indicating a limit cycle.}
\label{fig:oneoverrsquare_lc}
\end{figure}

Using Eqs.~\eqref{eq:p-estimate} and \eqref{eq:nonlinear-fit} we can now estimate the number
$P(s,\alpha)$ when given some pair $(s,\alpha)$. For instance, Fig.~\ref{fig:oneoverrsquare_lc}
contains the coupling constant $H_1(R_*,\pi/2)$ for the $1/r^2$ potential with displacement
$s=1/8$, for which we count $P(1/8,\pi/2)=7$ poles. In comparison, the estimate based on our
fit model computes the log-periodicity $\log(\lambda_\text{fit})(1/8, \pi/2) = 1.817\pm 0.117$
and the quotient $12/\log(\lambda_\text{fit})(1/8, \pi/2)=6.604\pm 0.425$ from which it predicts
\begin{equation}
P\left(\frac{1}{8},\frac{\pi}{2}\right) = 7^{+1}_{-0}.
\end{equation}

Having understood the behavior of $\log(\lambda)(s,\alpha)$, we can retrospectively legitimate the
normalization we agreed on in Sec.~\ref{sec:limit-cycle-loss}. In order to achieve $P\approx 100$ poles,
the mean of all pairwise log-pole differences must take values $\mu(\Lambda)\approx 1.2\times 10^{-2}$.
If the given potential is the $1/r^2$ potential or at least close to being LC, this corresponds to the
log-periodicity $\log(\lambda)(s,\alpha)$. When training the ensemble $\mathcal{S}$, we did not
have the $1/r^2$ potential in mind, yet, which is why we need to assure that it provides reasonable
displacements in that special case. Inserting the displacement $s=0.742\pm 0.029$ found in
Eq.~\eqref{eq:linear-fit} and the hyperangle $\alpha=\pi/2$ into Eq.~\eqref{eq:nonlinear-fit},
we obtain the log-periodicity $\log(\lambda_\text{fit})(s,\alpha)=0.130\pm 0.023$. This allows us
to estimate $P=93^{+16}_{-17}$, which agrees with the desired number of $P=100$ poles at the $1\sigma$ level.

The function $\log(\lambda_\text{fit})(s,\alpha)$ from Eq.~\eqref{eq:nonlinear-fit} used to model the
log-periodicity $\log(\lambda_\text{fit})(s,\alpha)$ also reproduces the expected limits
$\lim_{s\to-\infty}\log(\lambda_\text{fit})(s,\alpha) = \infty$, $\lim_{s\to\infty}\log(\lambda_\text{fit})(s,\alpha)
= 0$ and $\lim_{\alpha\to 0}\log(\lambda_\text{fit})(s,\alpha) = 0$, that are all beyond the fitting regime.
To what extend the obtained function is suitable for extrapolation is beyond the scope of this work
and reserved for future research.

\section{Discussion}
\label{sec:discussion}
In this paper we pursue an exploratory approach to identify LC potentials among a larger set of
discretized, attractive and singular two-body potentials based on unsupervised feature learning.
Here, the expression LC refers to two-body potentials whose coupling constant's RG flow satisfies an
RG limit cycle. The coupling constants $H_n(R_*,\alpha)$ themselves result from a delta-shell
regularization of the given potential $U\in\mathbb{R}^F$ and not only depend on a cutoff hyperradius
$R_*>0$, but also on some hyperangle $\alpha\in[0,\pi/2]$ and node index $n\in\mathbb{N}$. 
In contrast to the standard formulation of the Efimov effect, a delta-shell regulator that non-trivially 
depends on hyperspherical coordinates via $H_n(R_*,\alpha)$ is required in order to match 
the logarithmic derivatives of the Faddeev wavefunctions $\psi_{0,n}(R_*,\alpha)$ for arbitrary hyperangles. 
The $\psi_{0,n}(R_*,\alpha)$, again, are obtained by connecting local solutions of the low-energy Faddeev
equation via a generalized transfer matrix method. 
Unsupervisedly training a boosted ensemble $\mathcal{A}$ of ${C=8}$ convolutional VAEs $\mathcal{A}_i:
\mathbb{R}^F\to\mathbb{R}^L$, $i=1,\ldots,C$, to reconstruct the scaled potentials $\widetilde{U}$
from the training set $X\subset \mathbb{R}^F$ yields a low dimensional representation of all relevant
discretized two-body potentials and allows to relegate the search for LC potentials from the high-dimensional
feature space $\mathbb{R}^F$ to the lower-dimensional, effective latent space $\mathbb{R}^{C\times L}$.
The hierarchical structure of $\mathcal{A}$ is inherited by the ensemble-encoded potentials we
refer to as latent curves $\zeta\in\mathbb{R}^{C\times L}$. Each latent curve can be understood as
the ordered sequence of encoded contributions $\bm{\zeta}_i$ of all members $\mathcal{A}_i$. Later,
it is shown that for each member $\mathcal{A}_i$, the encoded contributions of training
potentials approximately form a standard distribution in $\mathbb{R}^L$. This is why latent curves
themselves and, consequently, synthetic potentials are simply drawn from a multivariate
standard distribution $\mathcal{N}(0, 1)$ in $\mathbb{R}^{C\times L}$.
As a measure for LC-ness we introduce the LCLoss $\mathcal{L}_\text{LC}$ that corresponds to the
quotient of the standard deviation and mean of the distribution $\Lambda$ of all log-pole differences
$p_{i+1}-p_i$ for any coupling constant $H_1(R_*,\pi/2)$ evaluated over the interval ${\log(R_*)
\in[-12-\log(\sqrt{2}), -\log(\sqrt{2})]}$ at $\alpha=\pi/2$ for the node index $n=1$. Leveraging
the fact that the shape of $\Lambda$ is invariant under the displacements $s$ provided by the
boosted ensemble $\mathcal{S}$, we normalize synthetic potentials such that the pole number of
their coupling constant takes values sufficiently close to $P\approx 100$. Thereby, sampling
errors within the calculation of the LCLoss are reduced.

Finally, we apply an elitist GA to fifty independent populations of latent curves, drawn from a
multivariate standard distribution $\mathcal{N}(0, 1)$ in $\mathbb{R}^{C\times L}$, and extract
the fittest individual, that is the latent curve with the lowest LCLoss, when completing the
final generation. Via ensemble-decoding, this distribution of fittest latent curves $\zeta\in
\mathbb{R}^{C\times L}$ implies a distribution of scaled potentials $\widetilde{U}=\mathcal{D}(\zeta)$
in feature space. The fact that these scaled potentials do not fall into several clusters, but behave
notably similar indicates that they accumulate around exactly one LC potential. Since the training
set covers a wide range of different singular potentials, we can safely dismiss the existence of
further LC potentials. Successfully fitting a model $a \log(R_*) + s$ to each of these scaled potentials
suggests the inverse square potential to be the desired, unique LC potential. By evaluating the LCLoss
of $1/r^p$ potentials with exponents close to $p=2$ not only for $\alpha=\pi/2$, but for smaller
hyperangles as well, we convince ourselves that the inverse square potential, indeed, minimizes
the LCLoss independent of the hyperangle. Thereafter, we study how the log-periodicity of the coupling
constant $H_1(R_*,\alpha)$ depends on the displacement $s$ and the hyperangle $\alpha$. Finding
a hyper-exponential dependence on $s$, we exemplarily demonstrate how this fit model can be
used to correctly predict the number $P$ of poles of the coupling constant.

In the context of RG limit cycles, the inverse square potential carries a special significance and
has been covered extensively in the literature. Hence, at first sight it is not surprising that the
result of our search is at least to some degree related to the inverse square potential. 
However, it is remarkable that there turns out to be
one unique LC potential that, in addition, is exactly the inverse square potential. Most interestingly,
the corresponding three-body system of identical bosons with the resulting pairwise inverse square
interactions is the same system whose three-body spectra have already been derived in
Ref.~\cite{PhysRevLett.108.213202}. Therefore, this paper can also be understood as a supplement
to Ref.~\cite{PhysRevLett.108.213202} highlighting the considered three-body system in view of
the behavior under a delta-shell regularization.

It is important to note that while investigating RG flows of coupling constants, cutoff hyperradii 
have only been continuously increased. This successive elimination of short-ranged degrees of freedom
uniquely renders the found limit cycle for the inverse square potential to be an infrared limit cycle.
Of course, further attention also needs to be paid to the discrete set of transition radii and the finite
range of all considered two-body potentials. These not only act as ultraviolet and, respectively, 
infrared regulators themselves, but also restrict all meaningful analyses of coupling constants to
a finite hyperradial interval. In constrast to the classical definition of a limit cycle, log-periodicity
of the coupling constant in this finite interval, therefore, already poses a sufficient criterion for being LC.

The non-convex and non-smooth loss surface topology of the LCLoss as defined in this paper suggests
to explore the latent space using GAs. At this point, it is important to emphasize that the existence
of an alternative LCLoss that is suitable for gradient descent based optimization while reliably
distinguishing between LC and non-LC potentials, seems plausible. However, a formulation of such
an alternative LCLoss requires deeper understanding of the fundamental mechanisms controlling the
RG flow of the considered coupling constant, which are yet to be uncovered by future research.

For further investigations it would be of great interest to determine the extent to which our
results can be generalized to more complex few-body systems. At this, special attention needs to
be paid to whether the found LC potential again corresponds to the two-body inverse square
potential and if the found solution is unique. If for whatever reason several LC potentials
should arise for some few-body system, it would be promising to compare the corresponding
bound state spectra and RG flows with each other.

\begin{acknowledgments}
We  acknowledge  partial  financial support by the Deutsche Forschungsgemeinschaft
(DFG, German Research Foundation) and the NSFC through the funds provided  to  the  Sino-German  Collaborative  Research  Center  TRR110  ``Symmetries  and  the  Emergence  of  Structure in  QCD''  (DFG  Project  ID
196253076  -  TRR  110,  NSFCGrant  No.  12070131001).  Support  was also provided
by the Chinese Academy of Sciences (CAS) President’s International Fellowship Initiative (PIFI)
(Grant No. 2018DM0034), by  Volkswagen Stiftung  (Grant  No.  93562),  and  by  the  EU Horizon 2020 (Grant No. 824093). Further, this project has received funding from the European Research Council (ERC) under the
European Union’s Horizon 2020 research and innovation programme (grant agreement No. 101018170).
\end{acknowledgments}

\appendix

\section{Commutation relation for $\mathcal{D}(R)$ and $\mathcal{I}$}
\label{sec:commutator}
The differential operator $\mathcal{D}(R)$ and the integral operator $\mathcal{I}$ as defined in Eqs.~\eqref{eq:operatord} and \eqref{eq:operatori} commute. Firstly, the only relevant contribution of $\mathcal{D}(R)$ to this commutator comes from the second-order hyperangular derivative. This reduces the commutator to 
\begin{equation}
\left[\mathcal{D}(R), \mathcal{I}\right] = \left[\frac{\partial^2}{\partial \alpha^2}, \mathcal{I}\right].
\label{eq:simplercommutator}
\end{equation}\\
We naively apply this second-order derivative to $\mathcal{I}\phi(\alpha)$ with $\phi$ being some hyperangular function and observe that we can, in fact, pull it into the integral, 
\begin{align*}
\frac{\partial^2}{\partial \alpha^2}\mathcal{I} \phi(\alpha) &= \frac{\partial^2}{\partial \alpha^2}\int_{\left|\nicefrac{\pi}{3}-\alpha\right|}^{\nicefrac{\pi}{2}-\left|\nicefrac{\pi}{6}-\alpha\right|} \! \mathrm{d}\alpha^\prime \, \phi(\alpha^\prime)\\
&= \mathrm{sign}\left(\frac{\pi}{6}-\alpha\right)\frac{\partial}{\partial \alpha}\phi\left(\frac{\pi}{2}-\left|\frac{\pi}{6}-\alpha\right|\right) \\
&\ \ \ +\mathrm{sign}\left(\frac{\pi}{3}-\alpha\right)\frac{\partial}{\partial \alpha}\phi\left(\left|\frac{\pi}{3}-\alpha\right|\right) \\
&=\phi^\prime\left(\frac{\pi}{2}-\left|\frac{\pi}{6}-\alpha\right|\right)-\phi^\prime\left(\left|\frac{\pi}{3}-\alpha\right|\right) \\
&= \int_{\left|\nicefrac{\pi}{3}-\alpha\right|}^{\nicefrac{\pi}{2}-\left|\nicefrac{\pi}{6}-\alpha\right|} \! \mathrm{d}\alpha^\prime \, \frac{\partial^2}{\partial {\alpha^\prime}^2}\phi(\alpha^\prime)\\
&= \mathcal{I}\frac{\partial^2}{\partial \alpha^2}\phi(\alpha).
\end{align*}
Using Eq.~\eqref{eq:simplercommutator} this, indeed, implies a vanishing commutator as well as simultaneous eigenstates,
\begin{equation}
\left[\mathcal{D}(R), \mathcal{I}\right] = 0.
\end{equation}

\section{Spectrum of $\mathcal{I}$}
\label{sec:eigenvaluei}
The equation of motion for the hyperangular sector of the low-energy Faddeev equation is a homogenous integro-differential equation that can be expressed via a differential and an integral operator $\mathcal{D}(R)$ and $\mathcal{I}$, respectively.
In App.~\ref{sec:commutator} we show that these operators commute, which implies the existence of simultaneous eigenstates. This finding has motivated us to start with considering the eigenvalue equation for $\mathcal{D}(R)$ instead of solving the hyperangular sector directly. Thereby, we identify its eigenstates as simple modes $\phi_n(\alpha)=\sin(2n\alpha)$ and intend to allocate these onto the whole equation Eq.~\eqref{eq:hyperangulareomsimple}. 

To further legitimize this approach, we must study how $\mathcal{I}$ acts on $\phi_n$. A simple evaluation of the integral and application of trigonometric addition theorems yields
\begin{align}
2n\ \mathcal{I}\phi_n(\alpha) &= 2n\int_{\left|\nicefrac{\pi}{3}-\alpha\right|}^{\nicefrac{\pi}{2}-\left|\nicefrac{\pi}{6}-\alpha\right|} \! \mathrm{d}\alpha^\prime \, \sin(2n\alpha^\prime)\\
&= \cos\left(2n\left|\frac{\pi}{3}-\alpha\right|\right)\notag \\
&\hphantom{=}\ -\cos\left(2n\left(\frac{\pi}{2}-\left|\frac{\pi}{6}-\alpha\right|\right)\right) \\
&= \cos\left(\frac{2\pi n}{3}\right)\cos\left(2n\alpha\right)\notag \\ 
&\hphantom{=}\ + \sin\left(\frac{2\pi n}{3}\right)\sin\left(2n\alpha\right) \notag \\
&\hphantom{=}\ -\cos\left(\pi n\right)\cos\left(\frac{\pi n}{3}\right)\cos\left(2n\alpha\right)\notag \\ 
&\hphantom{=}\ -\cos\left(\pi n\right)\sin\left(\frac{\pi n}{3}\right)\sin\left(2n\alpha\right)\label{eq:trigonometric1}
\end{align}
The first summand cancels the third summand due to
\begin{equation}
\cos\left(\frac{2\pi n}{3}\right) = \cos\left(\pi n\right)\cos\left(\frac{\pi n}{3}\right)
\end{equation}
such that all terms in Eq.~\eqref{eq:trigonometric1} proportional to $\cos(2n\alpha)$ vanish. Vice-versa, both remaining terms are proportional to $\sin(2n\alpha)$, which we already have identified as the eigenstates $\phi_n$. At this point, we can already tell that for each node index $n=1,2,3,\ldots$ $\phi_n$ is also an eigenstate of the operator $\mathcal{I}$. The corresponding eigenvalues can then be easily extracted from the eigenvalue equation,
\begin{align}
\mathcal{I}\phi_n(\alpha) &= \frac{1}{2n}\left[\sin\left(\frac{2\pi n}{3}\right)-\cos\left(\pi n\right)\sin\left(\frac{\pi n}{3}\right)\right]\phi_n(\alpha)\\
&= \frac{1}{n}\sin\left(\frac{2\pi n}{3}\right) \phi_n(\alpha).
\end{align}

\section{Logarithmic derivatives of the hyperradial wavefunction close to the delta-shell}
\label{sec:logarithmichyperradialderivatives}
In order to determine the coupling constant $H_n(R_*,\alpha)$, we have integrated the low-energy Faddeev equation oven infinitesimal interval containing the cutoff hyperradius $R_*$. As shown in Eq.~\eqref{eq:difflogderivs} this involves the hyperradial logarithmic derivative of the hyperradial zero-energy wavefunction immediately in front of and behind the delta-shell. Their calculation shall be explained here in greater detail.

\subsection{Outside the delta-shell}
The hyperradial zero-energy wavefunction for the regularized potential and the unregularized one need to match each other outside the delta shell. Close to the cutoff hyperradius they are given by\clearpage
\begin{widetext}
\begin{equation}
\begin{aligned}
\lim_{R\to R_*^+}f_{0,n}^{(i)}(R,\alpha) = A_{0,n}^{(i)}(\alpha)\sqrt{R_*}J_n[k_n^{(i)}R_*]+B_{0,n}^{(i)}(\alpha)\sqrt{R_*}Y_n[k_n^{(i)}R_*]
\end{aligned}
\label{eq:outsidedelta-2}
\end{equation}
Let $\Omega_n$ denote either the first-kind $J_n$ or second-kind Bessel function $Y_n$. The hyperradial derivative of $\sqrt{R_*}\Omega_n[k_n^{(i)}R_*]$ is evaluated as follows: At first simply applying the product rule yields
\begin{equation}
\begin{aligned}
2\sqrt{R_*}\frac{\partial}{\partial R_*} \left(\sqrt{R_*}\Omega_n[k_n^{(i)}R_*]\right) &= \Omega_n[k_n^{(i)}R_*] + k_n^{(i)}R_*\left(\Omega_{n-1}[k_n^{(i)}R_*]-\Omega_{n+1}[k_n^{(i)}R_*]\right)\\
&=2k_n^{(i)}R_* \Omega_{n-1}[k_n^{(i)}R_*] + \Omega_n[k_n^{(i)}R_*]-k_n^{(i)}R_*\left(\Omega_{n-1}[k_n^{(i)}R_*]+\Omega_{n+1}[k_n^{(i)}R_*]\right).
\end{aligned}
\label{eq:outsidedelta-1}
\end{equation}
We use the recursion relation 
\begin{equation}
k_n^{(i)}R_*(\Omega_{n-1}[k_n^{(i)}R_*]+\Omega_{n+1}[k_n^{(i)}R_*])=2n\Omega_n[k_n^{(i)}R_*]
\end{equation}

\noindent
for Bessel functions to express the last summand of Eq.~\eqref{eq:outsidedelta-1} in terms of $\Omega_n$. The resulting derivative 
\begin{equation}
\frac{\partial}{\partial R_*} \left(\sqrt{R_*}\Omega_n[k_n^{(i)}R_*]\right) = \frac{1}{2\sqrt{R_*}}\left(2k_n^{(i)}R_* \Omega_{n-1}[k_n^{(i)}R_*] + (1-2n)\Omega_n[k_n^{(i)}R_*]\right)
\label{eq:besselderivative}
\end{equation}
is involved when differentiating Eq.~\eqref{eq:outsidedelta-2}:
\begin{equation}
\begin{aligned}
&2\sqrt{R_*}\lim_{R\to R_*^+}f_{0,n}^{(i)\prime}(R,\alpha) \\&=A_{0,n}^{(i)}(\alpha)\left(2k_n^{(i)}R_* J_{n-1}[k_n^{(i)}R_*] + (1-2n)J_n[k_n^{(i)}R_*]\right) +B_{0,n}^{(i)}(\alpha)\left(2k_n^{(i)}R_* Y_{n-1}[k_n^{(i)}R_*] + (1-2n)Y_n[k_n^{(i)}R_*]\right)\\
&= 2k_n^{(i)}R_*\left(A_{0,n}^{(i)}(\alpha)J_{n-1}[k_n^{(i)}R_*]+B_{0,n}^{(i)}(\alpha)Y_{n-1}[k_n^{(i)}R_*]\right) + (1-2n)\left(A_{0,n}^{(i)}(\alpha)J_{n}[k_n^{(i)}R_*]+B_{0,n}^{(i)}(\alpha)Y_{n}[k_n^{(i)}R_*]\right)
\end{aligned}
\label{eq:outsidedelta-3}
\end{equation}
Then the logarithmic derivative of the hyperradial zero-energy wavefunction is given as the quotient of Eqs.~\eqref{eq:outsidedelta-2} and \eqref{eq:outsidedelta-3},
\begin{equation}
\lim_{R\to R_*^+}\frac{f_{0,n}^{(i)\prime}(R)}{f_{0,n}^{(i)}(R)}=k_n^{(i)}\frac{A_{0,n}^{(i)}(\alpha)J_{n-1}[k_n^{(i)}R_*]+B_{0,n}^{(i)}(\alpha)Y_{n-1}[k_n^{(i)}R_*]}{A_{0,n}^{(i)}(\alpha)J_{n}[k_n^{(i)}R_*]+B_{0,n}^{(i)}(\alpha)Y_{n}[k_n^{(i)}R_*]} + \frac{1-2n}{2R_*}.
\end{equation}
\end{widetext}
\subsection{Inside the delta-shell}
Due to $\widehat{u}_1=0$, the scenraio of taking the limit $R\to R_*$ from the inside is more complicated. For the modified momentum within the delta-shell, this implies 
\begin{equation}
\widehat{k}_n^{(1)}=k=0,
\end{equation}
since we consider the zero-energy wavefunction. Therefore, the limit $R\to R_*^-$ translates into a limit $k\to 0$ inside of the hyperradial zero-energy wavefunction itself,
\begin{equation}
\lim_{R\to R_*^-}\widehat{f}_{0,n}^{(1)}(R) = \sqrt{R_*}\lim_{k\to 0} J_n[kR_*],
\label{eq:deltainside-1}
\end{equation}
as well as for its derivative. Here we again make use of Eq.~\eqref{eq:besselderivative} and obtain
\begin{equation}
\begin{aligned}
&\lim_{R\to R_*^-}\widehat{f}_{0,n}^{(1)\prime}(R) \\&= \frac{1}{2\sqrt{R_*}}\lim_{k\to 0}\left( 2kR_*J_{n-1}[kR_*] + (1-2n)J_n[kR_*] \right).
\end{aligned}
\label{eq:deltainside-2}
\end{equation}\\

\noindent
Similar to the previous section where the limit $R\to R_*^+$ from the outside was taken, the logarithmic derivative follows as the quotient of Eqs.~\eqref{eq:deltainside-1} and \eqref{eq:deltainside-2},
\begin{equation}
\lim_{R\to R_*^-}\frac{\widehat{f}_{0,n}^{(1)\prime}(R)}{\widehat{f}_{0,n}^{(1)}(R)} = \lim_{k=0}\frac{k J_{n-1}[kR_*]}{J_n[kR_*]} + \frac{1-2n}{2R_*}.
\label{eq:deltainside-3}
\end{equation}
The Bessel functions of first kind, $J_n$, behave for small arguments $0<kR_*\ll \sqrt{n+1}$ like
\begin{equation}
J_n[kR_*] \approx \frac{1}{\Gamma(n+1)}\left(\frac{kR_*}{2}\right)^n.
\label{eq:besselasymptotic}
\end{equation}
Inserting Eq.~\eqref{eq:besselasymptotic} into Eq.~\eqref{eq:deltainside-3} eliminates the limit $k\to 0$ and finally yields
\begin{equation}
\lim_{R\to R_*^-}\frac{\widehat{f}_{0,n}^{(1)\prime}(R)}{\widehat{f}_{0,n}^{(1)}(R)} = \lim_{k=0}\frac{\frac{k}{\Gamma(n)}\left(\frac{kR_*}{2}\right)^{n-1}}{\frac{1}{\Gamma(n+1)}\left(\frac{kR_*}{2}\right)^n} + \frac{1-2n}{2R_*}=\frac{1+2n}{2R_*}.
\end{equation} 

\end{document}